\documentclass[acmsmall, authorversion]{acmart}

\AtBeginDocument{%
  }

  \author{Patrick Song}
  \email{pcsong33@gmail.com}
\affiliation{
  \institution{Harvard University}
  \city{Cambridge}
  \country{USA}
}
\author{Jayshree Sarathy}
\email{j.sarathy@northeastern.edu}
\affiliation{
  \institution{Northeastern University}
  \city{Boston}
  \country{USA}
}
\author{Michael Shoemate}
\email{shoematem@seas.harvard.edu}
\affiliation{
  \institution{Harvard University}
  \city{Cambridge}
  \country{USA}
}
\author{Salil Vadhan}
\email{salil_vadhan@harvard.edu}
\affiliation{
  \institution{Harvard University}
  \city{Cambridge}
  \country{USA}
}

\setcopyright{acmlicensed}
\copyrightyear{2024}
\acmJournal{PACMHCI}
\acmYear{2024} \acmVolume{8} \acmNumber{CSCW2} \acmArticle{472} 
\acmPrice{15.00} \acmMonth{11}\acmDOI{10.1145/3687011}

\received{January 2024} \received[revised]{April 2024} \received[accepted]{May 2024}

\acmISBN{978-1-4503-XXXX-X/18/06}
\usepackage[section]{placeins}
\usepackage{listings}
\usepackage{ctable}
\usepackage{tabularx}
\usepackage{tabularray}
\usepackage{enumitem}
\usepackage{etoolbox}
\usepackage{subcaption}
\AtBeginEnvironment{table}{%
\setlist[itemize]{nosep,
                  leftmargin=*,
                  before=\vspace{-0.6\baselineskip},
                  after=\vspace{-\baselineskip}
                  }
                            }
\usepackage{ragged2e}
\usepackage[normalem]{ulem}
\usepackage[font=itshape,vskip=0.75em]{quoting}
\usepackage[normalem]{ulem}
\usepackage{xcolor}
\newcommand{\eps}{\varepsilon}

\newif\ifshowcomments
\ifshowcomments
\newcommand{\PS}[1]{$\ll$\textsf{\color{orange} \small PS : #1}$\gg$}
\newcommand{\JS}[1]{$\ll$\textsf{\color{purple} \small JS : #1}$\gg$}
\definecolor{green(munsell)}{rgb}{0.04, 0.50, 0.12}
\newcommand{\addition}[1]{{\color{green(munsell)}#1}}
\newcommand{\removal}{\bgroup\markoverwith{\textcolor{red}{\rule[0.5ex]{2pt}{0.4pt}}}\ULon}
\else
\newcommand{\PS}[1]{}
\newcommand{\JS}[1]{}
\newcommand{\addition}[1]{#1}
\newcommand{\removal}[1]{}
\fi

\begin{document}
\title[Mental Models of DP Libraries]{
"I inherently just trust that it works": Investigating Mental Models of Open-Source Libraries for Differential Privacy}
% \JS{for first part of title, see if we can pull part of a quote}\PS{some options: Usability Hurdles to Security problems/Speaking the language of End Users}
%From theory to implementation: how open-source DP libraries shape mental models of privacy concepts

  % XXX is your article id# (for example: V8cscw052 = your article # is 52) \acmDOI{10.1145/XXXXXXX}
% \affiliation{%
%Draft comment features
% \affiliation{%

\renewcommand{\shortauthors}{Patrick Song, Jayshree Sarathy, Michael Shoemate, \& Salil Vadhan}

\begin{abstract}
% \JS{need to hook into a larger issue - open data? responsible AI? trustworthy infrastructures?}
% The push for open data requires rigorous and usable privacy tools to protect data subjects. 
Differential privacy (DP) is a promising framework for privacy-preserving data science, but recent studies have exposed challenges in bringing this theoretical framework for privacy into practice. These tensions are particularly salient in the context of open-source software libraries for DP data analysis, which are emerging tools to help data stewards and analysts build privacy-preserving data pipelines for their applications. While there has been significant investment into such libraries, we need further 
%there has been little 
inquiry into the role of these libraries in promoting understanding of and trust in DP, and in turn, the ways in which design 
of these open-source libraries can shed light on the challenges of creating trustworthy data infrastructures in practice.  

In this study, we use
qualitative methods and mental models approaches to analyze the differences between conceptual models used to design open-source DP libraries and mental models of DP held by users. Through a two-stage study design involving formative interviews with 5 developers of open-source DP libraries and user studies with 17 data analysts, we find  that DP libraries often struggle to bridge the gaps between developer and user mental models. In particular, we highlight the tension DP libraries face in maintaining rigorous DP implementations and facilitating user interaction. We conclude by 
offering practical recommendations for further development of DP libraries. 
\end{abstract}




\maketitle

\section{Introduction} 

Calls for open data have spurred new pipelines for sharing, accessing, and analyzing rich datasets. At the same time, initiatives around responsible AI call for rigorous privacy methods to protect the sensitive information of data subjects. The need for such privacy protections has been demonstrated through the long list of privacy attacks on seemingly anonymized datasets over the last few decades \cite{cohen22, sweeney00, sweeney02}.

One promising framework for privacy protection is \emph{differential privacy} (DP) \cite{Dwork06}, which involves adding noise to statistical releases to mask the contribution of any individual data point. DP offers provable privacy guarantees and mathematical quantification of the tradeoff between privacy and accuracy~\cite{wood2018differential,dwork2014algorithmic}. Over the last decade, DP has been deployed at a range of institutions, including Apple \cite{Tang17}, Google \cite{Erlingsson14}, Microsoft \cite{Ding17}, the US Census \cite{Abowd18}. 

To address the growing interest in deploying DP, 
institutions have invested in open-source software libraries for DP, which are emerging tools for data stewards, developers, practitioners, and analysts to study, test, and deploy DP analyses~\cite{OpenDP,Holohan19,GoogleDP}. The goal of these libraries is to promote correct use of, as well as trust in, DP. However, recent work studying the usability of these libraries has shown that not all DP tools fulfill these goals~\cite{ngong23}.

Studying open-source libraries for DP not only sheds light on making DP more usable, but also provides insights into key tensions around building trustworthy data pipelines in practice. User interactions with open-source DP libraries reveal misalignments between user mental models of DP with conceptual models exposed by the library software. 
More broadly, open-source libraries for DP are important sites to examine the challenges of 
%enacting responsible (and in particular, privacy-preserving) AI in practice as well as the tensions involved in 
retooling data science pipelines towards new ethical standards~\cite{Sarathy23}. 

In this study, we 
%bring together perspectives from HCI and usable privacy to 
analyze the programmatic frameworks of open-source DP libraries from the perspectives of library designers and data analysts. First, through interviews with developers of DP libraries, we explore the design choices developers make in implementing theoretical DP concepts in a practical setting and highlight the tradeoffs libraries make between usability and functionality. Second, using the findings from the developer interviews, we design a user study to analyze the challenges and complexities data practitioners face when using DP libraries.
Our study participants are data analysts and researchers who have interest in integrating privacy into their workflows, but have little experience programming DP algorithms. Analyzing this subset of users gives insight into the gaps between mental models of users and the conceptual model of open-source DP libraries. We also investigate the potential of open-source libraries to reinforce fundamental DP concepts, such as \emph{sensitivity} and \emph{the privacy-loss budget}. Specifically, we consider the following research questions.
\begin{quote}
\textbf{RQ1:} To what extent do user mental models of DP align with conceptual models and design of DP libraries?
\\[5 pt]
\noindent \textbf{RQ2:} What are usability challenges of DP libraries?
\\[5 pt]
\textbf{RQ3:} How do DP libraries communicate notions of trustworthiness and correct implementation to users?
\end{quote}
Through these questions, we aim to understand the role of DP libraries in empowering data analysts, and to provide suggestions on bridging the current gaps in usability and trust.

Our study consisted of semi-structured formative interviews with 5 DP library developers and a user study with 17 data analysts. Both the formative interviews and user studies included a mental model diagram-drawing exercise, designed to elucidate the understanding of DP libraries, across a variety of expertise levels. For the user study, we investigated two prominent DP libraries, OpenDP \cite{OpenDP} and Diffprivlib \cite{Holohan19}, exploring how well participants understood DP concepts and completed tasks. Our contributions include the following:
\begin{itemize}
\item Based on the formative interviews, we provide insights from the perspective of DP library developers and highlight areas of particular interest for the user study, including: 
(1) aligning DP library implementations with data science conventions, (2) facilitating intuitive workflows for data analysts, while also rigorously maintaining privacy guarantees and encouraging privacy-preserving data practices, and (3) understanding how users conceptualize notions of trust and security when using DP libraries.

\item From the user study, we find that comparing developers' conceptual models with users' mental models elucidates crucial gaps between theory and practice of DP libraries. 
We also trace how participants' mental models change as they interact with
documentation  
and error and warning messages.
We find that the DP libraries tested (Diffprivlib and OpenDP) do not effectively bridge misalignments between mental and conceptual models,
especially for participants with less DP experience. Yet, these less experienced participants exhibited the most change in their mental models of DP after using the libraries.
Finally, our study highlights that trust in DP libraries is multifactorial, stemming not only from open-source code but also 
usability of the programming framework, the clarity of documentation, and the external support from institutions. 

\item Finally, we offer practical recommendations for the development of open-source DP libraries, including:
(1) using expressive error messaging that better communicates potentially unintuitive, DP-specific privacy violations, (2) designing libraries to speak the language of the end user by highlighting accuracy, in addition to privacy, (3) incorporating visualizations to convey the probabilistic nature of DP, and (4) establishing norms around data analysis that align with the practical constraints of DP.
\end{itemize}
This paper is organized as follows. In Section~\ref{sec:background}, we introduce differential privacy and contextualize the investigation of the usability of DP libraries and understanding of DP concepts through mental models. Next, in Section~\ref{sec:formative-interviews}, we present findings from interviews conducted with DP library developers. Motivated by the discussions and insights from these interviews, we describe the design and results of a user study with data analysts using a DP library in Sections~\ref{sec:comparing-libs},~\ref{sec:user-study} and~\ref{sec:findings}. Finally, in Section~\ref{sec:discussion}, we contextualize our findings within broader goals of building trustworthy data pipelines and provide practical recommendations for improving the usability of DP libraries.

\section{Background and Related Work}
\label{sec:background}
We provide some background on differential privacy and situate our work within recent literature on usable differential privacy and privacy communication. We also define the concept of mental models and provide motivations for using mental models as a way to analyze user understanding of DP libraries. 
\subsection{Differential Privacy}
Differential privacy (DP), introduced by Dwork, McSherry, Nissim,
and Smith in 2006 \cite{Dwork06}, is a mathematical definition of privacy that measures how much information a mechanism for making a statistical release reveals about any one individual in the dataset. Informally, a DP mechanism $\mathcal{M}$ produces similar distributions of outputs on similar datasets, which means that the outputs of the mechanism will reveal little about any individual's data. ``Similarity of datasets'' is captured by the concept of neighboring datasets, which are defined as a pair of datasets that differ in the value of at most one individual. 

More concretely, a mechanism $\mathcal{M}$ satisfies $\eps-$DP if, for every pair of neighboring datasets $D \sim D'$ and any set of outputs $S$, the ratio of the probabilities $\mathcal{M}(D) \in S$ and $\mathcal{M}(D') \in S$ is bounded by $e^{\eps}$. The constant $\eps$ is known as the privacy-loss parameter. A smaller value of $\eps$ provides a stronger privacy guarantee, while a larger one provides a weaker guarantee. The formal definition is written as follows:
\begin{definition}[$\varepsilon$-DP \cite{Dwork06}] Let $\mathcal{M}: \mathcal{X}^n \rightarrow \mathbb{R}$ be a randomized mechanism. $\mathcal{M}$ is $\varepsilon$-\emph{DP} if for every pair of neighboring datasets $D\sim D'$and every subset $S \subseteq \mathbb{R}$:
$$
\Pr[{\mathcal{M}}(D) \in S] \leq e^{\epsilon} \cdot \Pr[\mathcal{M}(D') \in S]
$$
where the probabilities are taken over the randomness of $\mathcal{M}$.
\end{definition}

Given its rigorous mathematical guarantees and rich theoretical literature, DP has become a ``gold-standard'' for privacy-preserving data analysis. Across industry, non-profit groups, and academia, organizations have implemented and deployed data releases using DP. DP has enabled organizations to publicly release datasets while providing formal privacy guarantees to its data subjects. For example, the Wikimedia foundation used DP to publish statistics about how many distinct users visited each Wikipedia page on each day, from each country \cite{adeleye23}. The US Broadband Coverage dataset, released by Microsoft, quantifies the percentage of users who have access to high-speed Internet across the US \cite{pereira21}. Both Google and Facebook used DP to quantify mobility metrics for their users during the COVID-19 pandemic \cite{Meta, aktay20}. Notably, DP was used in the production of several of the data products in the 2020 Decennial US census \cite{Abowd18}. Across all of these deployments, practitioners continually noted that existing tools and software were not enough to meet the needs of each data release scenario \cite{boyd2022Differential, Hawes20, Drechsler23}. This gap has motivated the development of usable, open-source software tools that support a wide range of DP applications, which we will discuss below.

\subsubsection{Open-Source DP Software}
\label{sec:OSS}
Over the past few years, several open-source DP applications have been developed with the goal of communicating relevant DP concepts and supporting decision-making around DP releases. These include interfaces such as DP Creator \cite{gaboardi16}, DPComp \cite{hay16}, DPP \cite{john2021decision}, and Overlook \cite{thaker20}, which all attempt to portray accuracy and/or risk implications of the privacy loss parameter $\eps$ in order to support more informed privacy decisions. 
Additionally, over the past few years, several organizations have developed DP libraries in order to help data practitioners implement DP functions. These include Diffprivlib~\cite{Holohan19}, OpenDP~\cite{OpenDP}, Tumult Analytics \cite{Hay22}, and GoogleDP~\cite{GoogleDP}. We briefly describe these libraries here.

\addition{
\begin{itemize}
    \item \textit{Diffprivlib} \cite{Holohan19}: A general-purpose library for experimenting with differential privacy developed by IBM. It prioritizes simplicity and accessibility for users, with functionality mirroring that of existing data science libraries, like \texttt{NumPy} and \texttt{scikit-learn}.
    \item \textit{OpenDP} \cite{OpenDP}: A modular collection of statistical algorithms that utilizes two kinds of operators used in existing DP programming frameworks: \emph{measurements} and \emph{transformations}. OpenDP also incorporates privacy maps, chaining, and input and output domains as core programming elements.
    \item \textit{Tumult Analytics} \cite{Hay22}: A higher-level interface built upon Tumult Core, a privacy foundation library inspired by the same programming framework motivating OpenDP. Tumult Analytics gives users the ability to release aggregate information from sensitive datasets with DP, and uses an interface similar to the data science libraries \texttt{PySpark} and \texttt{Pandas}.
    \item \textit{GoogleDP} \cite{GoogleDP}: A suite of tools including both ``building block” libraries written in C++, Go, and Java, as well as higher-level frameworks. These frameworks include tools like Privacy on Beam, an end-to-end differential privacy framework built on top of Apache Beam, DP Auditorium, a library for auditing differential privacy guarantees, and a differential privacy accounting library, used for tracking privacy budget.
\end{itemize}
}
%\JS{use bullet points for readability. Expand if possible}
\removal{Diffprivlib \cite{diffprivlib}, developed by IBM, is a general-purpose library for experimenting with differential privacy developed by IBM..  The OpenDP library \cite{OpenDP}, on the other hand, is a modular collection of statistical algorithms that utilizes two kinds of operators used in existing DP programming frameworks: \emph{measurements} and \emph{transformations}. OpenDP also incorporates privacy maps, chaining, and input and output domains as core programming elements. Tumult Analytics \cite{Hay22} is a higher-level interface built upon Tumult Core, a privacy foundation library inspired by the same programming framework motivating OpenDP. Tumult Analytics gives users the ability to release aggregate information from sensitive datasets with DP, and uses an interface similar to the data science libraries \texttt{PySpark} and \texttt{Pandas}. GoogleDP \cite{GoogleDP} is a suite of tools including both ``building block” libraries and higher-level frameworks (Privacy on Beam, command line interface for ZetaSQL).}

\addition{In the first stage of our study (Section~\ref{sec:formative-interviews}), we interrogate the conceptual models of all four of these libraries. Based on the formative interviews, we narrow our focus in the second stage of the study (Section~\ref{sec:user-study}) to the two libraries that exhibit starkly contrasting conceptual models: OpenDP and Diffprivlib. In Section~\ref{sec:comparing-libs}, we provide an in-depth comparison of these two libraries and discuss how their differences can elucidate challenges of trust and usability for DP software.} 

\subsection{Usable differential privacy}
\addition{In software development, usability is often measured through a system's ability to be easily learned, understood, and operated \cite{Kitchenham05}. Given that users of DP software are often tasked handling sensitive data and protecting user privacy, we emphasize that usable DP systems should not only be easy to operate, but should also clearly define to users correct and accurate implementation patterns. Thus, in the context of our study, we extend our definition of \textit{usable DP} to include the notion of correct implementation, in addition to frictionless interactions and ease-of-use.}

Across the various deployments of DP, practitioners have documented numerous challenges around making this technically complex framework usable \cite{ngong23, Zhang18}. Correspondingly, usable DP has become a rapidly growing area of interest for privacy, security, and HCI researchers. We discuss three burgeoning lines of research: (1) trustworthiness of DP, including implementation vulnerabilities and governance issues, (2) communication of DP's guarantees to users, and (3) usability challenges.

\subsubsection{Trustworthiness and Safety.} First, 
%trust in DP's privacy protections has been shown to be a tricky issue. Even when DP tools aim to be rigorous about their guarantees, 
researchers and practitioners have faced many difficulties when translating DP from theoretical literature to usable software. Several works have shown how easy it is to implement DP algorithms incorrectly, creating vulnerabilities that make deployments susceptible to adversarial attacks and data leaks \cite{Cascuberta22, Mironov12, lokna2023, Kifer11}. These vulnerabilities are particularly important to address given that DP is considered, and described as, a mathematically provable way to protect data releases.
In addition, Smart et al. \cite{Smart22} and Sarathy \cite{Sarathy21} describe the `privacy theater' of DP, where lack of transparency into parameters used in DP to control privacy-utility tradeoff may mislead data subjects about protections offered. Smart et al. show that DP’s guarantees are unintuitive to data subjects, and that subjects rely on other factors, such as trust in data collectors, more than actual privacy protections when making data sharing decisions~\cite{Smart22}.
Thus, trust and governance have become important open questions for DP's continued use~\cite{Agrawal21,Dwork2019}.
\subsubsection{Communicating DP guarantees.} Second, researchers and practitioners have identified communication as a key open problem for DP. 
Cummings, Xiong, and Smart show that data subjects who are non-experts in DP struggle to make sense of the privacy protections offered \cite{cummings21, xiong20, Smart22}. Xiong, for example, investigate reasons why data subjects decide to share their data, showing that explanations of DP increase willingness to share data even while comprehension of these explanations remains low.
Karegar, Nanayakkara, Franzen, develop and test methods to communicate probabilistic DP guarantees to data subjects \cite{Karegar22, nanayakkara22, Franzen22}. The latter two works create quantified explanations for privacy loss parameter epsilon, which has proved challenging for data curators to set and communicate. 
Recent studies have also focused on visualization tools that can be used to convey the notion of privacy loss to data practitioners \cite{Panavas23, nanayakkara22} .

\subsubsection{DP interfaces and software.}
Finally, building usable tools for DP remains a challenge for researchers and practitioners. Sarathy et al. \cite{Sarathy21} investigate the DP Creator interface developed by OpenDP~\cite{OpenDP}, finding that the constraints imposed by the interface conflict with fundamental practices of data science, creating barriers for use by data curators and analysts.
 Garrido et al. \cite{garrido22} highlight an ‘academic-industrial DP utilization gap’ and specific technical barriers that DP tools must address, yet remain optimistic about the promise of DP for streamlining data processes and creating meaningful pathways for data exploration.

With the exception of two studies, there has been little research into the usability of DP libraries and how libraries communicate trust to data analysts. First, Zhang et al. \cite{Zhang22} evaluated the performance of OpenDP, Google DP, Pytorch Opacus, Tensorflow Privacy, and Diffprivlib on statistical queries and machine learning models. While this study compared the accuracy and utility of different libraries on different statistical operations, the study focused on performance metrics like memory loss and utility, rather than usability. Second, concurrent work by Ngong et al. \cite{ngong23} evaluated the usability of four DP tools: OpenDP, Diffprivlib, Pipeline DP, and Tumult Analytics. While Ngong et al. used several usability metrics in order to evaluate and compare DP libraries, understanding how DP libraries communicate trustworthiness and correct implementation was not a main focus for their study. 

In the next section, we outline how mental models can be used as a framework to understand questions surrounding usability and trust for DP libraries.

\subsection{Mental Models}
The recent emergence of open-source DP libraries developed by several different organizations provides an opportunity to compare the different programmatic approaches taken by DP developers and their effects on mental models of users. Given the difficulties of translating theoretical DP concepts to applied settings \cite{Ashena21, Dwork2019}, there is a need to evaluate the extent to which software successfully implements DP concepts. 
In this study, we use conceptual and mental models as a framework to interpret both user understanding of DP concepts and usability of DP library software. 

A \emph{mental model} is defined as an internal representation of external reality that reflects an individual's understanding of how a system works or behaves \cite{craik43, johnson83}. HCI researchers frequently rely on mental models to represent human interaction with systems~\cite{Katzeff94,Landriscina2013}. 
In user-interface design, these user mental models are often compared against \emph{conceptual models} of systems, which are representations or descriptions of a system. Previous work has emphasized the importance of exposing a software's conceptual model to the user so that their mental model aligns with the conceptual model \cite{Jackson21}. Analyzing user interactions with computer systems through conceptual and mental models can elucidate the cognitive organization of technical concepts and  dynamic interactions between them \cite{Furlough18, Story99, Gero20}. 
\addition{\subsubsection{Mental Models for Complex Concepts}
Although DP is a complex framework, it is still valuable to study using mental and conceptual models across various levels of prior knowledge. Prior work has studied people’s understanding of general complex systems, highlighting the nuances that can be captured using diagramming approaches and emphasizing differences between mental models of experts and non-experts~\cite{bravo2010bridging}. For example, studies  have shown that non-experts’ mental models focus on concrete components and singular or centralized points of causality, whereas experts center abstract elements and recognize multiple, decentralized avenues of action~\cite{hmelo2004comparing,jacobson2001problem,resnick1998diving}. These prior works indicate that using mental model analysis is both appropriate and valuable in our study’s context.}

\subsubsection{Mental Models in Privacy Research}
Mental models have been employed in various research settings as a framework in which to understand user concerns, knowledge, and expectations of privacy systems. Coopamootoo et al. \cite{Coopamootoo14} applied the technique of cognitive mapping to depict user mental models of privacy, illustrating the relationship between privacy concern and behavior. In an interview-based study, Wang et al. \cite{wang2023} used an interactive digital board to elucidate mental models of content moderation reporting behavior on end-to-end encrypted  messaging platforms. Mental models have also been used to highlight user misconceptions of privacy concepts. Through a qualitative study involving a drawing task-based lab study, Puadel et al. \cite{Paudel23} highlighted participant misunderstandings of third-party applications’ data collection and sharing practices. Dumaru et al. \cite{Dumaru23} conducted a qualitative study that depicted the relationship between user mental models of the internet and misconceptions of security tools.

\subsubsection{Mental Models in Open-Source DP Libraries}
There has been limited related work applying mental models to exploring open-source libraries. This includes Storey et al.'s work \cite{Storey99} that explored graphical representations of software as a way of creating more intuitive connections between software elements and program source code and documentation. More recently, researchers found that the addition of dataflow graphs to data pipelines in the machine learning library TensorFlow aided in user understanding and debugging of models \cite{Wongsuphasawat18}. Other work examines mental models of users learning Apache Beam, an open-source distributed data processing API, in order to inform design modifications~\cite{Horvath19}. 
We believe that mental model approaches are particularly important for understanding the usability of DP, and how well it maintains its rigorous protections when put in the hands of non-expert data practitioners.

In the next section, we describe a set of formative interviews designed to elucidate the conceptual model of DP libraries. We later compare these with mental models of users described in Section \ref{sec:models} to understand gaps between design and use of DP libraries.

\section{Phase 1: Formative Interviews with Developers}
\label{sec:formative-interviews}
We began by conducting a set of formative interviews with library developers to better understand the challenges for usability of DP software. While several proprietary DP software tools exist, we focused these interviews on \emph{open-source} libraries, as we were interested in understanding how transparency and nuanced privacy concepts are communicated through open-source software. In addition, we hoped to elucidate the conceptual models that library developers have of open-source DP libraries. 

\textbf{Recruitment.}
We recruited 5 participants through direct email requests, relying on internal connections formed through an online community of DP researchers and practitioners. We targeted individuals who work in organizations across academia, industry, and research. Our participants included research scientists, software developers, and program managers who have worked on or are currently working on open-source DP library development at Google, Tumult Labs, OpenDP, and IBM. \addition{A brief description of each of these libraries is provided in Section \ref{sec:OSS}. Given the small sample for this formative study, we do not include additional details about which participant was involved with which library.} 

\textbf{Interview Protocol.}
Each interview lasted about 60 minutes and was recorded with consent. These interviews were conducted in a semi-structured style \removal{based on}\addition{tailored to} the background and experience of the participant. We first asked developers a series of open-ended questions (see Section~\ref{sec:user-study-instructions} for a full list), regarding design choices they have made while implementing DP algorithms; challenges around usability, trust, and communication they have observed or experienced; and insights they have gleaned from developing open-source DP software.

Next, we showed developers examples of a mental model diagram from an internet security user study (see Figure~\ref{fig:modelexample}) to provide an understanding of what a mental model diagram looks like. Then, we asked developers to choose an example of a computation on data (eg. a summary statistic) and map out how their library implements a DP computation and executes it on a sensitive dataset. This exercise, conducted using the Zoom whiteboard tool, was designed to elucidate the developer's conceptual framework of a DP library. 

\textbf{Analysis.}
Using the notes, transcripts, and recordings from each interview, we conducted an inductive thematic analysis. Two study team members open-coded the data and through iterative discussions and collaboration with the entire team, we grouped the codes into themes, which are discussed below.

\textbf{Ethics.}
All parts of the study were approved by our institutional IRB. As the group of DP library developers is small and well-known within the DP community, we explained to our interview participants the risks of participating, encouraged them to decline answering any questions, and have removed identifying information when reporting results.

\textbf{Findings.}
\label{sec:formativeobservations} 
Based on our analysis, we identified four main issues raised by developers around design and usability of DP libraries. These issues included difficulties surrounding library usability and creating intuitive debugging processes for users, especially those with less formal DP experience. Developers also highlighted the importance of user trust in both DP and library implementations. We summarize these themes in the first two columns of Table~\ref{table:formative}, \addition{and expand on the findings below.} The third column of Table~\ref{table:formative} describes how we used each of these themes to guide the design of the tasks and questions in the second phase of the study, which is described in Section~\ref{sec:user-study}.

\begin{center}
\newcolumntype{s}{>{\hsize=.5\hsize}X}
\begin{table}[htp]
  \begin{tabularx}{\textwidth}{p{0.20\textwidth}p{0.38\textwidth}p{0.34\textwidth}}
    \FL
 \textbf{Theme} & \textbf{Description} & \textbf{Incorporation into DP Library User Study} 
    \ML
    Library Usability & Challenging to make a DP library usable and useful for a variety of different levels of DP expertise. & Included participants with a wide range of DP expertise, working across both academia and industry \addition{(Section~\ref{sec:participants})}.
    \\ \NN
    Debugging Errors & Unclear how successfully users interpret errors output by DP libraries. & Included an analysis task using a perturbed dataset designed to encourage documentation exploration and debugging \addition{(Section~\ref{sec:perturbation})}. 
    \\ \NN
    Communicating Trust &  Trust and correct implementation are communicated in many ways through DP libraries, to varying degrees of success.  & Asked participants about the role DP libraries play in communicating trust during the mental model diagram exercise \addition{(Section~\ref{sec:mental-model-drawing})}. 
    \\ \NN
    Benefit of Open-Source & Open-source DP software benefits from a community that can vet current code and contribute to existing software. & Asked post-task question regarding the value of open-source code in regards to DP software \addition{(Section~\ref{sec:user-study-instructions})}.   %\JS{Use ref to correct section here}  \PS{corrected}  
    \LL
  \end{tabularx}
  \vspace{10pt}
    \caption{Four themes capturing important topics and issues faced by DP library users, from the perspective of library developers. Each theme was incorporated into the DP library user study as described in the third column.}
    \label{table:formative}
\end{table}
\end{center}
\subsection{Misalignments between data science norms and DP implementations}
\label{sec:misalignments}
One of the main difficulties mentioned by DP developers related to the usability of libraries was the misalignment between data science conventions and DP implementations. These difficulties arise when a library's adherence to a particular theoretical framework causes the software interface to be unintuitive to users without prior DP knowledge. For example, OpenDP's framework of chaining transformations and measurements together is largely based on theoretical framework commonly used in building algorithms in research settings \cite{OpenDP}. However, P3 noted that many data scientists are more familiar with a workflow where they append operations to a dataset incrementally, making OpenDP's current workflow hard to use for some. Specifically, P3 mentioned the need to create DP tools from a top-down approach,
\begin{quoting}
I think there's often user confusion about what is a measurement versus a transformation. Why do I make one versus the other? We need to focus on mapping our core concepts onto things that the user has interacted with before. Instead of a bottom up approach, maybe a better way to to look at it is from the top down. What is the user trying to get done? And how do we facilitate that? (P3).
\end{quoting}

\begin{figure}
     \centering
     \begin{subfigure}[h]{0.75\textwidth}
         \centering
         \includegraphics[width=\textwidth]{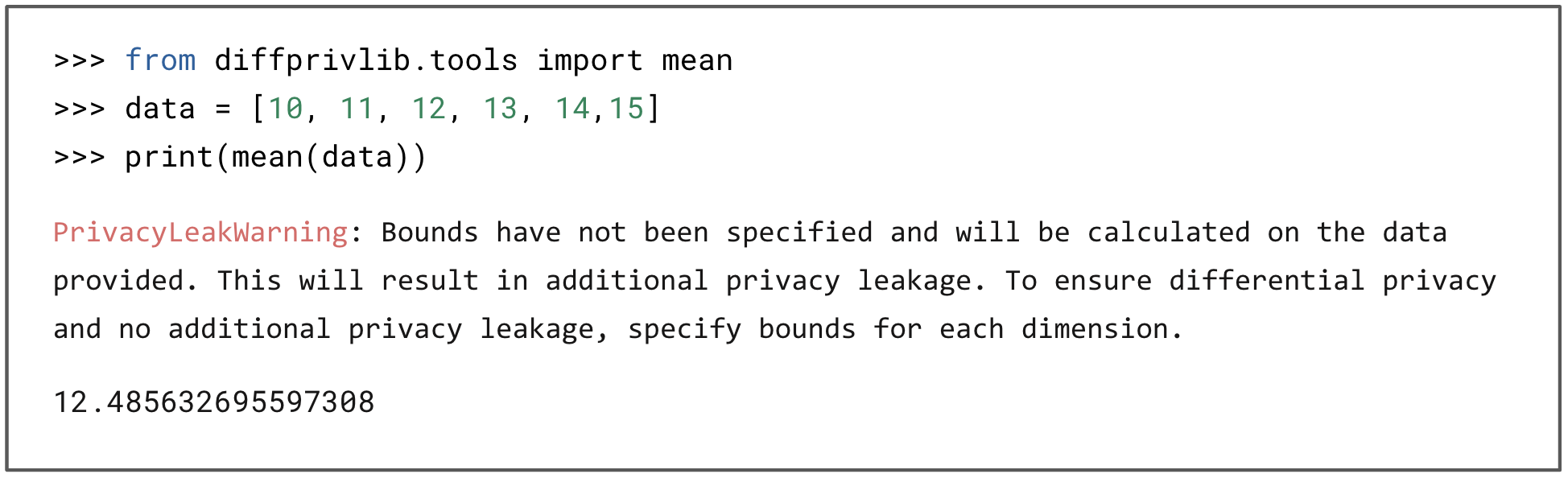}
         \caption{Diffprivlib Bounds Privacy Warning}
     \end{subfigure}
     \hfill
     \begin{subfigure}[h]{0.75\textwidth}
         \centering
         \includegraphics[width=\textwidth]{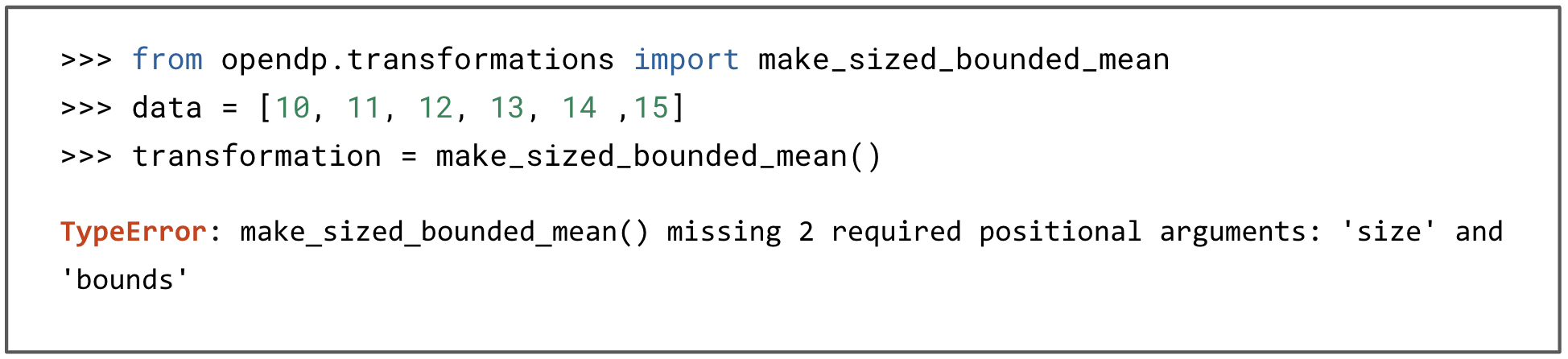}
         \caption{OpenDP Bounds Type Error}
     \end{subfigure}
     \hfill
     \caption{\addition{Comparison of Diffprivlib's privacy warning and OpenDP's error message when calculating a mean on a dataset without specifying bounds.}}
     \label{fig:bounds}
\end{figure}

In order to align more closely with existing mental models of data scientists, Diffprivlib provides similar functionality with existing data science libraries. Many of Diffprivlib's DP functions are similar in implementation to their non-private counterparts in NumPy and Scikit-learn, making the use of Diffprivlib more accessible to those who have familiarity with common data science open-source tools. While this approach may reduce flexibility for users who wish to build algorithms from scratch, it doesn't require library users to develop a strong mental model of privacy concepts. For Tumult Labs, their Tumult Analytics platform is a user-friendly interface built on top of their core library. Similarly, OpenDP is moving towards developing a Context API for their library, which doesn't require users to fully understand all the complexities of OpenDP's underlying framework. This movement from libraries highlights the inherent difficulties in fully exposing a library's programmatic framework to end users, and the need to abstract the parts of a library that are harder to understand.

\subsection{Tradeoff between privacy and ease-of-use in DP libraries}
\label{sec:tradeoff}
Our interviews highlighted the different approaches that libraries take to manage the tensions between maintaining privacy guarantees in a rigorous way versus prioritizing usability of the library for non-experts in DP. These different approaches contributed to the different expectations that library developers had in designing error messages to help facilitate user debugging and understanding of fundamental DP concepts. 

Multiple participants brought up the example of user-defined clamping bounds to illustrate this. In Tumult Analytics and OpenDP, if a user attempts to calculate a DP mean from a dataset without specifying clamping bounds, this will result in an error that forces users to input  bounds before proceeding with any further computations. In Diffprivlib, however, attempting to calculate a DP mean without inputting bounds will not result in an error. Instead, the library will compute a DP mean statistic using bounds calculated directly from the dataset, accompanied by a warning message that the computation may result in a privacy leakage (Figure \ref{fig:bounds}).

In discussing this example, the developers highlighted the nuances of these opposing design choices.
From the perspective of P2, while Diffprivlib's approach may compromise the DP guarantee, it is important for reducing user friction and encouraging further exploration of the library. On the other hand, P5 felt that compromising on privacy guarantees for the sake of usability can be problematic. At the same time, P5  highlighted additional difficulties that can arise from requiring users to input dataset bounds when they may not know how to do so.
\begin{quoting}
The data scientist who is not versed in DP will think: `how do I overcome the limitation? I'm going to privately look at what the min and max is in my data and then put that in the clamping bounds.' So these kind of usability hurdles are  indirectly also security problems because they can encourage people to do the wrong thing (P5).
\end{quoting}

\subsection{Communicating notions of trust and security} 
\label{sec:communicatingtrust}
Given that open-source DP software is increasingly being used in sensitive data workflows, many of the developers indicated the importance of communicating the correct implementation of privacy algorithms to users. Yet, the ways in which trust and security are emphasized to users varies across libraries. For OpenDP, a unique component of their approach is the integration of mathematical proofs into their code review and documentation process. Mathematical verification is one of the key differentiators of OpenDP, and and is one way in which privacy guarantees are communicated. However, P3 questioned this approach:
\begin{quoting}
I can see users being glad, knowing that the proofs are there, and that someone else has looked at them, because that means they don't need to. But the real, actual practical value of those proofs ... other than researchers, it could very well be that these proofs are not interesting to anyone then (P3). 
\end{quoting}
P3 maintained that proofs are only one component of OpenDP's approach to communicate correct implementation to users, and are part of the library's larger efforts to encourage discussion and contribution in the open-source community. Similar to OpenDP, Tumult Analytics is built using components from Tumult Core, a design structure that guarantees the proof of privacy for a DP program. Both developers from OpenDP and Tumult Labs maintained that the ability to derive privacy proofs directly from a DP program constructed from these lower-level building blocks is an integral part of the library's design framework.

Many of the developers noted the benefits of open-sourcing software in communicating privacy and security goals. Two of the developers noted that the open-source community around DP libraries has helped developers spot bugs and verify the correctness of their algorithms. However, one developer noted that the importance of open-source extends beyond correcting code, making an analogy to the cryptography space:
\begin{quoting}
For instance, in cryptography, it makes no sense to use proprietary algorithms that you hide the details of. You need transparency to build trust, especially in the privacy space (P5). 
\end{quoting}
In summary, these interviews highlighted the difficulties related to usability, error handling, and communicating trust and correct implementation that library developers feel users of DP libraries face.

\section{Highlighting Differences between OpenDP and DiffPrivLib} 
\label{sec:comparing-libs}
Using the findings from the formative study, we designed and ran a user study to examine how interacting in different DP libraries changed mental models and understanding of privacy concepts. The libraries we investigated were Diffprivlib and OpenDP, which were chosen due to their differing implementation of DP algorithms.

Created by researchers and developers from IBM, Diffprivlib \cite{Holohan19} is designed for researching, experimenting with, and creating differential privacy applications using Python. The library is designed around simplicity and ease of use, catering to a broad range of users. Diffprivlib leverages the functionality of the NumPy and Scikit-learn packages and many of Diffprivlib's private functions mirror the same parameter usage as their non-private counterparts in NumPy and Scikit-learn. OpenDP \cite{Hay20} is a community effort led by Harvard University to build trustworthy, open source software tools for statistical analysis of sensitive private data using differential privacy. OpenDP's core library, defines measurements and transformations as the two main operators in the library, similar to previous systems like PINQ \cite{mcsherry10} and Ektelo \cite{Zhang18}, and allows researchers or data scientists to make programmatic use of the library and its Python bindings. OpenDP's use of measurements and transformations, in addition to privacy maps and chaining, allows the library to enforce strict privacy practices and prevent operations that violate DP guarantees. For the purposes of this study, we examined OpenDP's Context API, a higher-level abstraction of OpenDP's core library.

\addition{Compared to other DP libraries, we felt that analyzing the combination of  Diffprivlib and OpenDP would give the greatest insight and learnings into how successfully users interact with two vastly different conceptual models of DP libraries.}   
For the user study, we expected Diffprivlib's conceptual model to be easier to grasp for those with experience using data-science libraries in Python, due to the library's similarity in functionality to NumPy and Scikit-learn. In contrast, given OpenDP's implementation of concepts from theoretical DP research, like measurements, transformations, and privacy maps, we expected the OpenDP's conceptual model to be familiar to those with more theoretical DP experience. In order to illuminate these differences, we explain in detail the following operations that we hoped to analyze in the user study in order to compare user mental models of OpenDP and Diffprivlib.

\subsection{Computing a DP Mean}
\label{sec:computingmean}

OpenDP requires users to bound the sensitivity of the mean query by first clamping the dataset using the $\texttt{clamp}$ method. Additionally, since the number of rows $n$ of the dataset is typically considered a private parameter, the $\texttt{resize}$ parameter establishes the dataset size (an assumption that can be made through public knowledge of the dataset), and imputes the value of extra rows as 0. The \texttt{laplace} method then infers the sensitivity of the mean query and adds the corresponding amount of laplace noise. Finally, the DP mean is computed using the \texttt{release} method (Fig \ref{fig:opendpmean}, see additional figures).

In contrast, Diffprivlib's DP mean implementation requires less code. The only required parameter for the \texttt{mean} function is the \texttt{array} parameter. If the \texttt{epsilon} and/or \texttt{bounds} parameters are not specified, Diffprivlib assumes an epsilon value of 1 and/or calculates the lower and upper bounds of the data directly from the array, respectively. Additionally, Diffprivlib assumes that the count of the dataset is public knowledge, and calculates the count directly from the dataset. While users can infer these assumptions from the arguments of the function, this implementation choice is not clearly defined in the documentation. These design choices are likely made in order to mirror the functionality of NumPy's mean function, which resembles Diffprivlib's argument definition in all but the \texttt{epsilon} and \texttt{bounds} parameters (Fig \ref{fig:diffprivlibmean}, see additional figures).

\subsection{Handling null values}
\label{sec:perturbation}

Both OpenDP and Diffprivlib provide functionality for accommodating null values in a dataset. In the user study, we ask participants to compute a DP mean on two datasets: a dataset where values greater than 45,000 are removed and a dataset where values greater than 45,000 are changed to null values. All other values less than or equal to 45,000 are equivalent across both datasets. 

OpenDP contains a \texttt{make\_impute\_constant} transformation which replaces all null values with a specified constant. On the other hand, Diffprivlib has a separate DP \texttt{nanmean} function, which behaves very similarly to its DP \texttt{mean} function, except that it ignores all null values when averaging up all of the values for computing the mean. However, when computing the sensitivity of the dataset, Diffprivlib uses the total count \textit{including the null values}. Thus, the amount of noise added when using the \texttt{nanmean} function on the dataset with null values is underestimated compared to when using the \texttt{mean} function on the dataset with removed values. We wanted to investigate how participants reacted to seeing this discrepancy during Task 1 of the programming portion of the user study (Figure \ref{fig:nanmean}, see additional figures).

\subsection{Privacy Accounting}
Lastly, we highlight the ways in which DP libraries mediate the use of a privacy budget. Privacy accountants in DP libraries typically encode the theoretical concept of sequential composition, where the total privacy cost of releasing multiple results of differentially private mechanisms on the same input data can be tracked and bounded. In OpenDP, the \texttt{Context} class handles multiple DP queries to the same dataset. The \texttt{Context} class requires five parameters and requires users to associate a specific dataset with a given privacy budget, defined by the \texttt{privacy\_unit} and \texttt{privacy\_loss} parameters. In contrast, Diffprivlib's budget accountant takes in four optional parameters: \texttt{epsilon}, \texttt{delta}, \texttt{slack}, and \texttt{spent\_budget}. Users don't need to specify the number of queries to be made upfront. Additionally, they can define the \texttt{spent\_budget} parameter in order to initialize the privacy budget with DP computations calculated previously. In both OpenDP an Diffprivlib, the library outputs an error when the privacy budget has been exhausted (Figure \ref{fig:budget}, see additional figures). 

\section{Phase 2: User Study Background and Methodology}
\label{sec:user-study}

In this section, we specify the study protocol for the DP library study, describing how the mental model diagram exercise and the data analysis tasks assigned to participants relate to the research questions we aim to investigate.

\subsection{Participants and Settings}
\label{sec:participants}
As a qualitative study whose aim is to be exploratory rather than generalizable, we aimed to recruit participants with  demonstrated interest in DP rather than seeking out data analysts interested in privacy more broadly. We did so for two reasons. First, those interested in DP are most likely to be the intended users of these tools. Second, this allowed us to work with participants who were invested in the development of DP tools and thus willing to be communicative while drawing mental models, looking through documentation, and completing tasks using DP software.

Therefore, all prospective participants were recruited through a preliminary questionnaire sent out on the OpenDP community mailing list. 
We chose to use this mailing list as it is the most comprehensive platform for reaching people interested in the practice of DP, and the only mailing list we know of that is not specific to a single DP library or deployment. Even though the list is organized by OpenDP, it contains developers and users of the other DP libraries (via OpenDP's larger goals of community-building across DP practitioners) as well as those who have never before used any DP software but are interested in learning more.
%\JS{any stats we can get on the makeup of the mailing list would be great}
\removal{While participants were primarily recruited through the OpenDP Community mailing list, most of this community does not consist of people who are already using OpenDP software, but just people who are interested in the practice of DP.}

On the questionnaire, participants were asked to fill out their familiarity level with Python and several Python libraries (SciPy, NumPy, Pandas, scikit-learn). They were also asked about any past experience they had with differential privacy, as well as any other data privacy methods or frameworks. \removal{In order to filter out people with extensive programming experience in the DP libraries in our study,} We also asked participants whether they had any prior programming experience with OpenDP or Diffprivlib. Our assignment of participants to use OpenDP or Diffprivlib in the user study was not random, but rather was done carefully based on their prior experiences and with an eye to the overall balance of our sample. Participants with prior programming experience in either OpenDP or Diffprivlib were assigned to the library they had not used before, or were removed from consideration for the user study if they had prior experience with both libraries. This was to ensure that our study could capture changes to participant's mental model that occurred via their first use of a library. Additionally, our assignment aimed to balance participant familiarity with DP across both libraries (ie. both sets of assignments contained a similar number of participants with low/high familiarity with DP).

In total, we recruited 17 participants (4 female, 13 male). 10 participants were research graduate students, and 7 participants were software developers. Of the 10 research students, 4 of them work directly with differential privacy, while the rest work in fields adjacent to differential privacy (statistics, machine learning, data visualization). All participants had a strong familiarity with Python. The studies were conducted virtually over Zoom software. Participants received a \$45 Amazon gift card as compensation for their time. The summary of the total participant sample is shown in Table \ref{table:participants}. 

\begin{table}
\begin{center}
\begin{tabular}{|c|c|c|c|}
    \hline
    Participant & Background & DP Familiarity Level \removal{Assigned Library}\\
    \hline
    \addition{O1} &  Researcher & Extremely familiar \removal{OpenDP}\\
    \addition{O2} & Researcher & Very familiar \removal{OpenDP}\\ 
    \addition{O3} &  Researcher & Very familiar \removal{Both}\\
    \addition{O4} & Researcher & Moderately familiar \removal{Both}\\ 
    \addition{O5} &  Developer & Very familiar \removal{OpenDP}\\ 
    \addition{O6} &  Developer & Very familiar \removal{OpenDP}\\ 
    \addition{O7} &  Developer &  Moderately familiar \removal{OpenDP}\\
    \addition{O8} &  Developer & Slightly familiar \removal{OpenDP}\\ 
    \addition{D9} &  Researcher & Extremely familiar \removal{Both}\\ 
    \addition{D10} &  Researcher & Very familiar \removal{Both}\\ 
    \addition{D11} &  Researcher & Very familiar \removal{Diffprivlib}\\
    \addition{D12} &  Researcher & Moderately familiar \removal{Diffprivlib}\\
    \addition{D13} & Researcher & Moderately familiar \removal{Diffprivlib}\\
    \addition{D14} &  Researcher & Moderately familiar \removal{Diffprivlib}\\
    \addition{D15} &  Developer & Moderately familiar \removal{Diffprivlib}\\
    \addition{D16} &  Developer & Slightly familiar \removal{Diffprivlib}\\
    \addition{D17} &  Developer & Slightly familiar \removal{Diffprivlib}\\
    \hline
\end{tabular}
\caption{Summary of user study participant sample. \addition{Participants with an O-prefix and D-prefix were assigned to OpenDP and Diffprivlib, respectively. Participants were assigned to libraries based on their prior experiences and with the aim of balancing familiarity with DP across both libraries (see Sec. \ref{sec:participants} for more details).}}
\label{table:participants}
\end{center}
\end{table}

\subsection{Study Protocol}
We conducted a 75-min study session with each participant, and with permission, recorded the session. Prior to the session, participants were asked to read a short non-technical primer on differential privacy. The participants were randomly assigned to complete tasks in either Diffprivlib or OpenDP for the length of the study. Due to the time and effort required to learn the syntax and framework of a DP library, we chose a ``between-subjects'' design for this study. \removal{However, a few of the participants agreed to extend the session and completed tasks in both Diffprivlib and OpenDP.} We opened each user session with background questions about the participant. Then, participants were guided through a set of activities, including a mental model drawing exercise, documentation walkout, and programming task. 

Interviews were recorded and transcribed, and in total, we collected about 25 hours of audio and video recordings. All parts of the study protocol were approved by an IRB.

\subsubsection{Mental model exercise}
\begin{figure}
    \centering
    \includegraphics[width=0.75\linewidth]{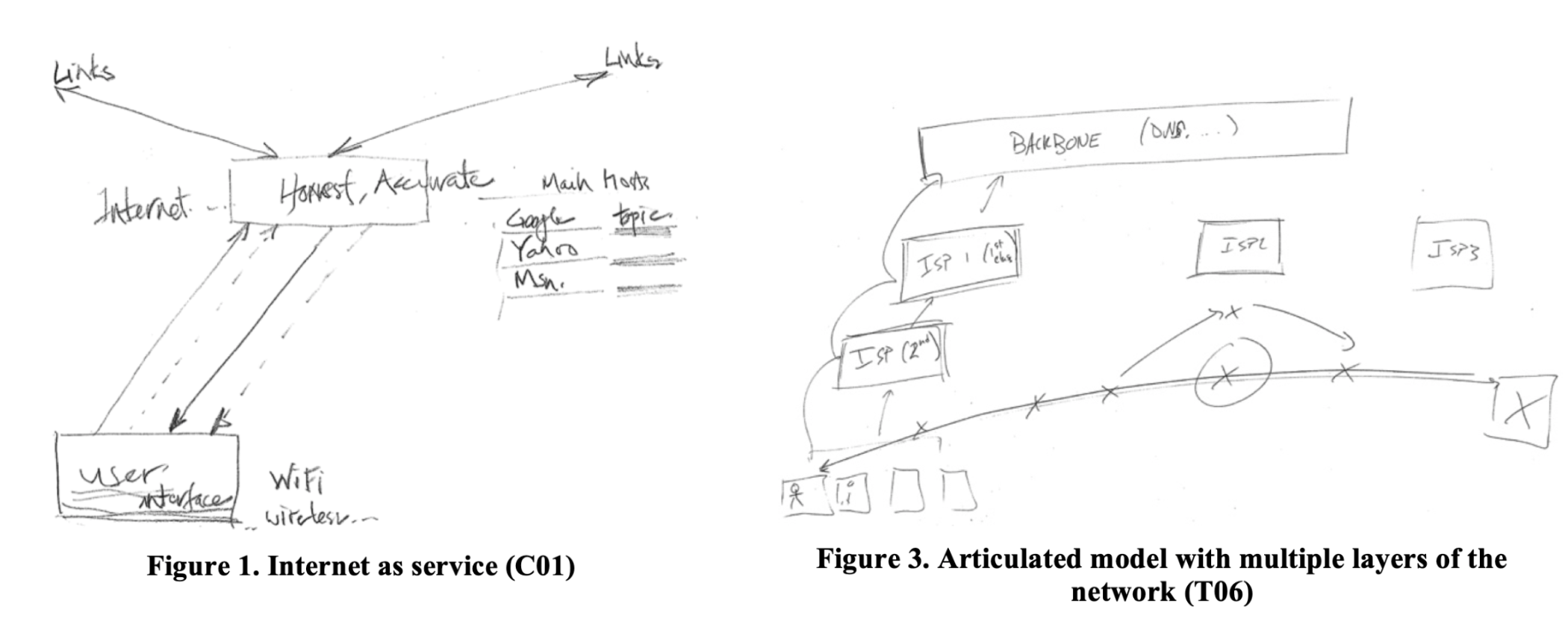}
    \caption{Example mental models shown to participants from internet security user study \cite{Kang15}}
    \label{fig:modelexample}
\end{figure}

Participants were guided through a mental model exercise similar to the one described in the formative interview section. We showed participants examples of a mental model diagram from an internet security user study (Figure \ref{fig:modelexample}) and once they understood what a mental model diagram looks like, we asked participants to choose an example of a computation on data (eg. a summary statistic) and map out how a DP library implements the computation and executes it on a sensitive dataset. We also asked participants to highlight the parts of the library that they thought would be most vulnerable to a privacy violation. Participants were instructed to verbalize their thought process as they drew, consistent with traditional think aloud protocols \cite{Ericsson80}. The diagramming was completed through the Zoom Whiteboard tool.
\addition{To address the impact of prior knowledge of DP, we saved these mental models and had participants update them after completing the task.}

\subsubsection{Documentation walkthrough}
After completing the drawing exercise, participants were asked to examine documentation of the libraries they were assigned to. Participants assigned to OpenDP were first asked to navigate to the OpenDP Welcome Page, which gives an overview of the components and capabilities of the library. Then, we asked them to read an overview of the OpenDP Programming Framework, a conceptual model that defines the characteristics of privacy-preserving operations and provides a programmatic way for components to be assembled into programs with desired behavior. Finally, we instructed them to read through an example notebook that explained the functionality of the Context API. Participants assigned to Diffprivlib were first asked to navigate to Diffprivlib's homepage on GitHub, where they read a brief introduction to the library outlined in the repository's README file. Then, they were instructed to go to the docs page of Diffprivlib, where they were directed to the Tools section of the page, which contains documentation for several general purpose DP functions. 

\subsubsection{Programming tasks}
\label{sec:tasks}
Afterwards, participants worked through two data analysis tasks using functions from their assigned DP library. These tasks were designed to highlight the differences in implementation of Diffprivlib and OpenDP, as discussed in Section \ref{sec:comparing-libs}. We describe the tasks below:
\begin{itemize}
    \item \textit{Task 1: DP Mean.} Participants were first asked to examine a mock dataset containing demographic information including income, location, and age. Then, they were instructed to calculate the DP mean of the income column in the dataset. Afterwards, we presented participants with two modified versions of the dataset, and asked them to compute the DP mean of these changed datasets. In the first version, all income values less than 45,000 were removed. In the second version, all income values less than 45,000 were changed to NaN values.
    \item \textit{Task 2: DP Histogram.} Upon completing the mean task, participants were then asked to compute a DP histogram distribution of the age values in the original mock dataset. Participants were again encouraged to verbalize their thought process, and ask questions if they were stuck. Throughout both tasks, participants were asked to track their privacy budget usage.
\end{itemize}
These tasks were completed in an interactive Jupyter notebook environment using the software tool Deepnote. After completion of the programming tasks, we closed the user study session by asking a few questions related to the usability of the library. We also asked participants if they would make any changes to the initial mental model diagram drawn towards the beginning of the session.

\subsubsection{Reflection and updating mental model.}
 At the end of the study session, participants were asked several open-ended questions to reflect on the task (See Section~\ref{sec:user-study-instructions}). In addition, participants were asked to update their initial mental model diagrams. This allowed us to gauge the evolution of users' engagement with DP concepts in a more exploratory, open-ended manner compared to testing users' comprehension via factual questions. In particular, by comparing the pre- and post-task versions of mental models, we were able to understand how the library impacted users’ mental models, across various levels of prior knowledge or experience with DP.

\subsection{Analysis}
The notes, transcripts, and audio/video recordings from each session were analyzed by two of the study team members using a reflexive thematic analysis approach \cite{Braun06}. Guided by our own positionality as DP researchers and the research questions, we engaged in a collaborative and iterative process of open-coding the data. After discussions with the entire study team, this resulted in 25 codes. These codes included: Importance of correct implementation, Usefulness of visualizations, Misunderstanding fundamental DP library norms; a full list is found in the Appendix ~\ref{sec:codebook}. These codes were combined into themes that we discuss in Section~\ref{sec:findings}.

\subsection{Ethics and positionality}
All of the study team members are DP researchers. One is additionally a developer of one of the DP libraries. Our analyses aimed not to promote one DP library over another, but to critically analyze the role of open-source DP libraries in creating trustworthy, privacy-preserving, and usable data pipelines. 

All parts of the study were approved by our institutional IRB. The study procedure used only publicly available datasets that are non-sensitive. We took care to ensure that participants were comfortable declining to answer any questions and made sure to remove any identifying information when reporting our results.

\begin{center}
\newcolumntype{s}{>{\hsize=.5\hsize}X}
\renewcommand{\arraystretch}{1.5}
\begin{table}[htp]
  \begin{tabularx}{\textwidth}{p{0.30\textwidth}p{0.65\textwidth}p{0.20\textwidth}}
    \FL
 \textbf{Research Question} & \textbf{Findings}
    \ML
    To what extent do conceptual models of DP libraries align with user
    mental models of DP? (RQ1) & 
    \begin{itemize}
        \item Those with less experience in DP displayed mental models of DP as a black box (Sec \ref{sec:simplemodels})
        \item More experienced participants described concepts such as privacy budget, mechanism design, and access controls in their mental models (Sec \ref{sec:articulatemodels})
        \item Compared to OpenDP, Diffprivlib's documentation more effectively bridged the library's conceptual model with users' mental models (Sec \ref{sec:accesibledoc})
    \end{itemize}
    \ML
    What are usability challenges of DP libraries? (RQ2)&
    \begin{itemize}
        \item Prioritizing rigorous DP implementations often interfered with task completion and usability (Sec \ref{sec:taskcompletion})
        \item Error and warning messages failed to correctly orient user mental models (Sec \ref{sec:errors})
        \item Libraries lacked consistent tooling to help users verify their DP results (Sec \ref{sec:verify})
    \end{itemize}
    \ML
    How do DP libraries communicate notions of trustworthiness and
    correct implementation to users? (RQ3) & 
    \begin{itemize}
        \item Documentation is an important yet limited source of conveying trust (Sec \ref{sec:documentation})
        \item Libraries balance maintaining an intuitive codebase with signalling correct implementation to users (Sec \ref{sec:usabilitytrust})
        \item Trust is multifactorial, relying on openness, proofs, and institutional contexts (Sec \ref{sec:open-source})
    \end{itemize}
 
    \LL
  \end{tabularx}
  \vspace{10pt}
    \caption{Summary of research questions and main findings}
    \label{table:findings}
\end{table}
\end{center}

\section{Findings}
\label{sec:findings}

\removal{Here, we discuss
the results of the study.} 
Our main, qualitative findings from the user study are organized according to the three research questions, and are summarized in Table~\ref{table:findings}.
\addition{Supplementary quantitative metrics regarding the tasks are provided in Table~\ref{table:quantitative}}. 
\addition{Overall,} we explain how participant mental models changed through their interaction with DP libraries, and how conceptions of trust and utility were received by participants. \removal{The connection between themes uncovered in the user study findings and the research questions are summarized in Table \ref{table:findings}.}

\subsection{Mental Models of DP Libraries}
\label{sec:models}
\begin{figure}
    \centering
    \begin{subfigure}{0.5\linewidth}
        \centering
        \includegraphics[width=\linewidth]{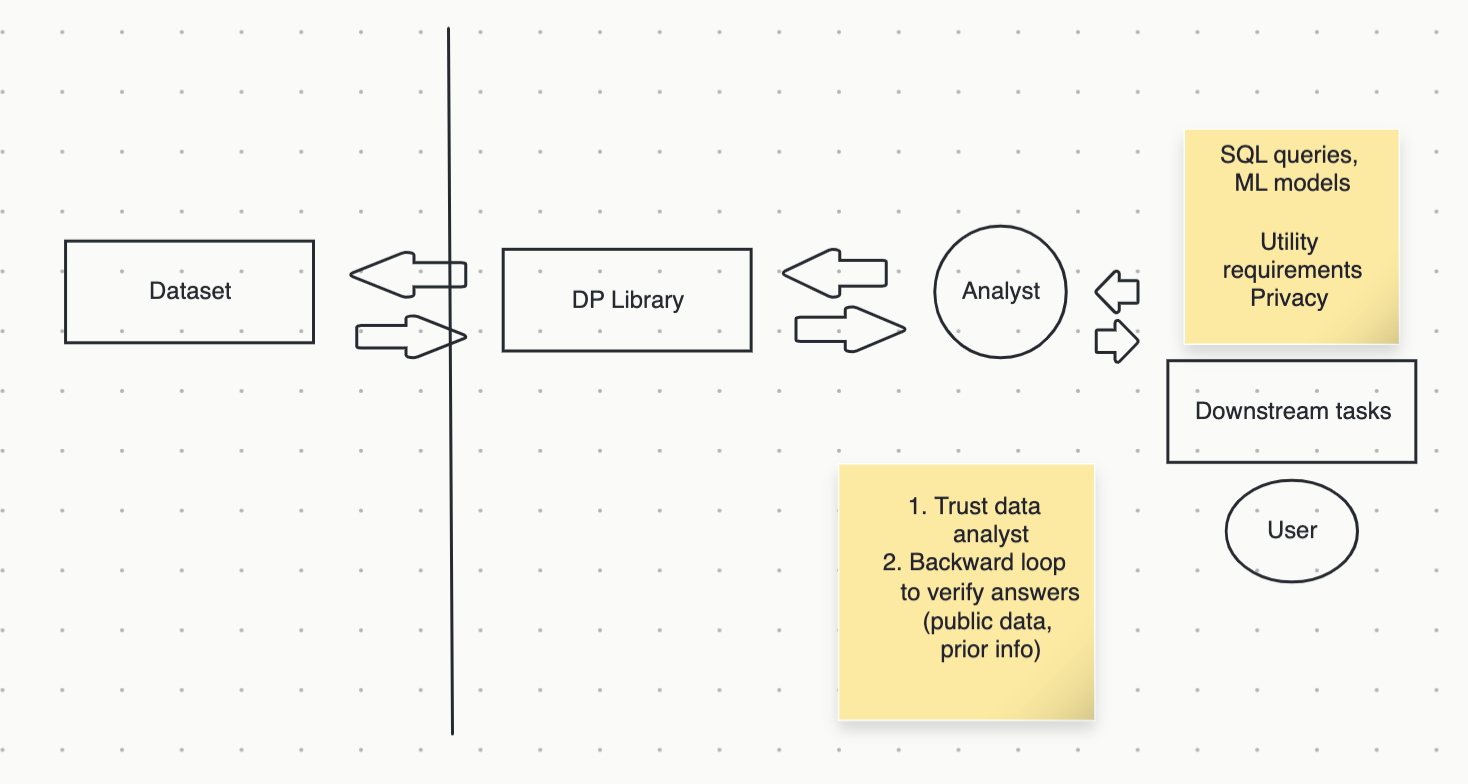}
        \caption{Bidirectional Complex Model (D10)}
        \label{subfig:b10model}
    \end{subfigure}
    \hfill % Add horizontal space between subfigures
    \begin{subfigure}{0.45\linewidth}
    \centering
    \includegraphics[width=0.9\linewidth]{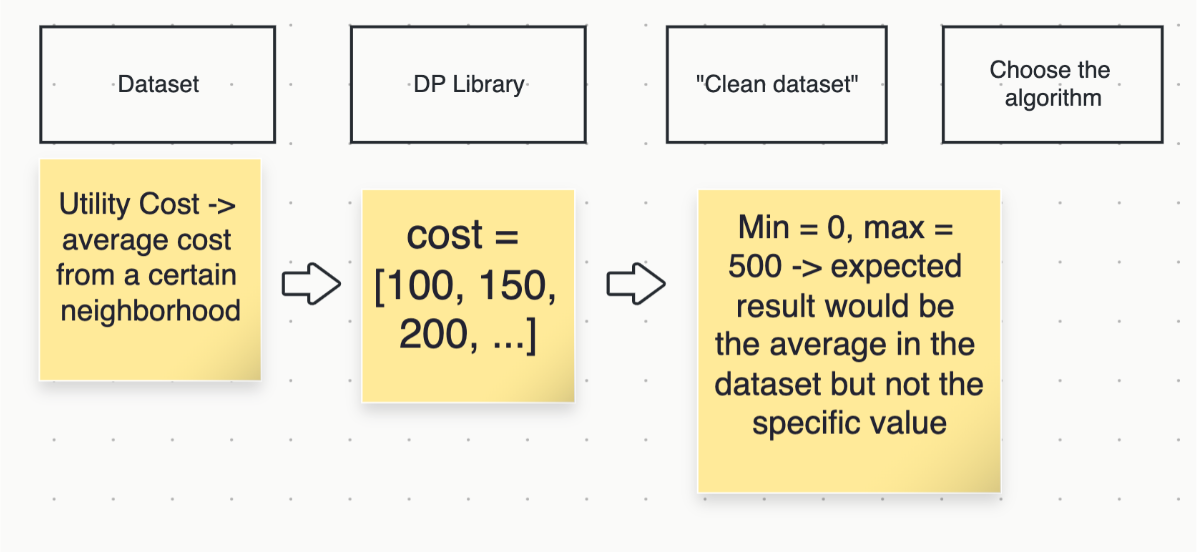}
    \caption{Utility Cost Computation (D17)}
    \label{subfig:b17model}
    \end{subfigure}
    \caption{Comparison of Complex and Simple Mental Models}
    \label{fig:comparison}
\end{figure}

In this section, we discuss how user mental models align with the conceptual framework of DP libraries (RQ1). Participants were first asked to complete a drawing exercise designed to elucidate their mental models of DP libraries. These mental models reflected participants' initial understanding of DP libraries prior to any examination of code documentation or completion of programming tasks. As expected, those with less DP experience tended to view DP libraries as simple mechanisms\removal{ (Figures \ref{fig:b17model} and \ref{fig:b14model}, see additional figures)}, while participants with more DP experience typically had models that were complex and detailed (Figure \ref{fig:comparison}). These mental model diagrams provided personalized baselines that participants would return to after perusing further details about DP implementations and use. 

Next, participants explored documentation of the DP library they were assigned to. Through having participants talk aloud as they explored documentation, we aimed to elucidate differences between their own mental models of DP libraries and the conceptual model of the DP library they were exploring. One of our main findings was that Diffprivlib's documentation was easy to understand for all participants, while for participants using OpenDP, the documentation led to some degree of confusion for participants who were less experienced in DP. This finding concurs with the results of Ngong et al. \cite{ngong23}, which found that OpenDP's documentation was much less intuitive for users than Diffprivlib's documentation.

\subsubsection{Those with less experience in DP displayed mental models of DP as a black
box}
\label{sec:simplemodels}
Out of the 7 developer participants without DP research experience, 6 of them constructed their diagrams as a simple, unidirectional (or non-directional) set of inputs and actions (Fig \ref{subfig:b17model}). Only 2 of these simple diagrams encoded any notion of privacy parameter selection or sensitivity bounding in their diagrams, fundamental concepts in DP. In general, simple mental model diagrams showed assumptions that a DP library should have an encompassing set of privatizing functions. However, these participants often did not describe how the library would produce these privatized outputs, and viewed the library as a black box. For example, one participant stated:
\begin{quoting}
I imagine I would run the privatizing function. My inputs would be the dataframe and some privacy parameters (O8).
\end{quoting}
Additionally, when constructing their mental model diagrams, developer participants often spoke of using differential privacy in a manner that wasn't fundamentally different from a normal data analysis process. 
\begin{quoting}
 I'd imagine you could do the process without privacy. And then, if you wanted to, you could insert whatever tools are used from the library into the process to make it private. (D10)
\end{quoting}
This view was also reflected in some of the mental model drawings. As shown in Figure \ref{fig:b14model} (see additional figures), O14 referred to the output of a DP library function as a ``DP-ed statistic'', but didn't specify the nuances behind creating such a statistic. Broadly speaking, while participants were able explain a hypothetical workflow in which DP was utilized, they lacked the knowledge to describe DP beyond a basic understanding of noise addition. 

\subsubsection{More experienced participants described concepts such as privacy budget,
mechanism design, and access controls in their mental models}
\label{sec:articulatemodels}
Researcher participants typically had more complex diagrams that encoded a greater number of technical complexities related to DP. While most participants' indicated some notion of a privacy-loss parameter, researcher participants included concepts like allocating a privacy budget, choosing a privacy mechanism, and mediating access to the sensitive dataset. Figure \ref{fig:A1model} (see additional figures) illustrates how one participant considered the relationship between accounting for privacy loss, choice of privacy mechanism, and bounding sensitivity when performing a differentially private linear regression. 

4 out of the 7 complex models incorporated a bidirectional flow of information, in which the data analyst both makes queries to and receives outputs from the DP library. One participant visualized this by drawing a set of arrows pointing in opposite directions in processes involving the dataset, DP library, and data analyst, reflecting the viewpoint that performing DP computations is an interactive process, rather than a one-way flow of operations (Figure \ref{fig:A1model}, see additional figures). 

In addition, 3 of the complex models also included some visual representation delineating the privatized outputs from the initial dataset. In Figure \ref{subfig:b10model}, this is represented by a vertical line separating the input dataset from the DP library operations. Another participant visualized this concept by crossing out an arrow drawn from the initial dataset to a DP linear regression output, indicating that the private output should not reveal information about the underlying dataset (Figure \ref{fig:A1model}). Related to this concept, one participant asked about the hypothetical diagramming situation:
\begin{quoting}
So I have a question before we start. Do I have access to the dataset? Am I able to look at the data? (D10)
\end{quoting}
This inquiry highlights a nuanced understanding that looking at the raw dataset can lead to privacy violations, one that was not brought up by participants without DP background. This speaks to a key source of uncertainty that can arise in a real-world situation when working with a sensitive dataset using DP software. 

There were a few similarities between the mental model drawing process for all participants. Nearly all participants either drew or verbalized some notion of a noisy or less accurate output, indicating knowledge of the privacy-utility tradeoff in DP. Additionally, all 10 researcher participants and 5 of the 7 developer participants either wrote or mentioned the privacy loss parameter epsilon when constructing their diagram. This suggests that non-experts may be exposed to these basic concepts of DP, but that exposure does not always translate to more complex mental models.

\addition{Participants were asked towards the end of the study whether they would change anything to their mental model drawings after reading the library documentation and completing the library tasks. Predictably, almost all of the participants who were more experienced in DP made no modifications to their mental model drawing. However, several participants who were less experienced in DP modified their diagrams by drawing additional arrows and boxes to indicate the importance of inputting data bounds when calling DP functions. We expand upon participant interactions with concepts like bounding the sensitivity of datasets in Section \ref{sec:errors}.}

\subsubsection{Compared to OpenDP, Diffprivlib’s documentation more effectively bridged the library’s conceptual model with users’ mental models}
\label{sec:accesibledoc}
Next, we discuss participants' perceptions when exploring DP library documentation. In general, users found that Diffprivlib's documentation was easy to follow. In particular, users who had used libraries like NumPy and scikit-learn found Diffprivlib's documentation particularly accessible. 
\begin{quoting}
This is pretty similar to NumPy. If someone knows NumPy, [Diffprivlib] is very clear to understand. (D16)

\noindent Seems easy enough to understand. (D15)

\noindent I think this library's implementations are very friendly. (D17)
\end{quoting}
In contrast, OpenDP's documentation was noticeably harder to understand for participants. Even participants with significant prior DP experience struggled to grasp the library's programmatic framework and conceptual model.
\begin{quoting}
Oh, man, it's hard to track the documentation. (O6)

\noindent I think the documentation is very confusing for me, at least for someone who hasn't done DP. (O7)
\end{quoting}
Two main components of OpenDP's programming framework, \emph{measurement}s and \emph{transformations}, were sources of confusion for users. Recall that measurements are defined as randomized mappings from a private dataset or value to an arbitrary output value that is safe to release, and transformations are defined as deterministic mappings from a private dataset to another private dataset or value in OpenDP. While terms like measurements and transformations had previously been used in DP literature, they are not standard terminology. Therefore, participants found the library's naming conventions unintuitive and are sometimes in conflict with prior mental associations with these concepts.  
\begin{quoting}
The word choice of measurements is confusing; why wouldn't it be private outputs or something? (O4)

\noindent Construct a measurement? Okay, the documentation is probably going to explain this. I guess measurement is anything like average or median. (O2)
\end{quoting}
Both O2 and O4 thought that the term measurement referred to a different statistical concept than its meaning within OpenDP, highlighting a disconnect between their mental model and the library's conceptual model. Even though OpenDP's Context API attempted to abstract some elements of the programming framework from the documentation and code, participants found it difficult to parse through the documentation and asked many more questions while completing the documentation exploration than with Dffprivlib. This suggests that while OpenDP's programming framework may facilitate greater flexibility and verifiability, elements of the framework can be a barrier to usability. Efforts to abstract OpenDP's programming framework remain a challenge, and further work is needed to make it accessible to new users.

\subsection{Usability Challenges of DP libraries}
\label{sec:programming}
After finishing the documentation overview, we asked participants to complete data analysis tasks in their assigned DP library. By observing how participants interacted with the libraries, and how participants reacted to libraries' attempts to redirect actions via error and warning handling, we traced the usability challenges that affect each library (RQ2). In particular, we discovered a tension between providing correct and rigorous DP implementations and \addition{frictionless library use.} \removal{library usability.} Additionally, we found that Diffprivlib's warning messaging led to overconfidence in the privacy guarantees of a particular DP implementation, while OpenDP's error handling hindered user interaction with the library. Furthermore, after completing the tasks, participants found it difficult to verify the correctness of their own code implementations. We discuss these findings in more detail below\addition{, and provide a summary of task metrics by library in Table~\ref{table:quantitative}.}

\begin{table}
\begin{center}
\newcolumntype{C}[1]{>{\centering\let\newline\\\arraybackslash\hspace{0pt}}m{#1}}
\begin{tabular}{|C{1.8cm}|C{3cm}|C{3cm}|C{3.2cm}|}
    \hline
     & Average (StdDev) \newline \# of Errors \newline Encountered & Average (StdDev) \newline Minutes Spent & Total \# of Correct Implementations \\
    \hline
    OpenDP \newline (N=8) &  7.3 (3.1) & 18.4 (6.9) & 8 \\ \hline
    Diffprivlib \newline (N=9) & 1.5 (0.7) & 7.5 (3.2) & 3\\ %Cecilia Documentation: 19:47, Mean: 12:43, Histogram: N/A, Errors/Exceptions: 4
    \hline
\end{tabular}
\caption{\addition{Summary statistics regarding Task 1, where participants were asked to calculate a DP mean. (See Section~\ref{sec:tasks} for more details). With Diffprivlib, compared to OpenDP, participants encountered fewer errors, spent less time, and yet achieved fewer correct implementations. Task 2 statistics not included here as several OpenDP participants did not get to it.}}
\label{table:quantitative}
\end{center}
\end{table}

\subsubsection{Prioritizing rigorous DP implementations often interfered with task completion}
%\removal{usability}}
\label{sec:taskcompletion}
Overall, participants assigned to Diffprivlib completed more tasks than those assigned to OpenDP. All 9 of the participants assigned to Diffprivlib were able to complete the first DP mean task, described in Section \ref{sec:computingmean}. Of these participants, 2 of them also completed the histogram task, and 4 of them completed all of the tasks in the notebook. In comparison, none of the 8 OpenDP users were able to progress past the first DP mean task. Additionally, two of the OpenDP participants required significant help in completing the mean task.

One of the biggest impediments to task completion for OpenDP users was the number of errors and warnings thrown by the library. On average, an OpenDP user encountered 7.3 errors throughout the programming session while a Diffprivlib user encountered 1.5 errors. Additionally, participants assigned to OpenDP spent significantly longer time on average during the documentation overview part of the study \addition{(Table \ref{table:quantitative})}, meaning they generally had less time to complete the programming tasks. These results suggest that Diffprivlib presented a more familiar and intuitive workflow than OpenDP and that OpenDP typically required a steeper learning curve for participants.

However, faster task completion did not necessarily lead to correct implementations of the tasks. For example, for the first mean task, only 3 out of the 9 Diffprivlib participants correctly implemented the task on their first attempt. The remaining 6 participants only realized their error after prompting by members of the study team. Although users of OpenDP required significantly longer amounts of time on average to complete the first mean task compared to users of Diffprivlib \addition{(Table \ref{table:quantitative})}, all of them were able to complete the task correctly on their first attempt. These findings highlight an inherent tension between \addition{frictionless library interaction} and ensuring rigorous, and correct implementations for users.
 
\subsubsection{Error and warning messages failed to correctly orient user mental models}
\label{sec:errors}
While Diffprivlib users were able to finish their code implementations fairly quickly, 5 out of the 9 Diffprivlib users encountered an \emph{unspecified bounds} message when completing the DP mean task. This occurred when participants attempted to calculate a DP mean without providing lower and upper bounds of the data they were computing on (Figure \ref{fig:bounds}). The warning message relayed that the sensitivity of the function was calculated via direct computation on the dataset, resulting in an arbitrary privacy leak. 

Based on the amount of DP experience participants had, their reactions to this warning varied greatly. 2 Diffprivlib participants failed to notice the bounds warning message until prompted by one of the members of the study team. Those who were less experienced in DP either misunderstood the warning or downplayed the potential significance of not specifying any bounds. 
\begin{quoting}
So again, if I'm just developing quickly, it's fine. It does do a pretty decent job, because the maximum is probably a pretty decent bound to set. (D13)
\end{quoting}
This quote highlighted a significant misunderstanding on the part of B13; the maximum is actually the best possible upper bound one can set in terms of utility. However, computing this value directly from the dataset is a privacy violation, as it reveals information about the underlying dataset \cite{Dwork10}. Another participant recognized the danger of Diffprivlib's approach, noting that outputting a warning instead of an error increases the likelihood that a user will simply ignore the message.
\begin{quoting}
As a DP practitioner, I know that [bounds are] important. But someone who is not that familiar with DP \ldots yeah, I think this warning might be skipped by people. (D10)
\end{quoting}
Those who were more experienced in DP were critical of Diffprivlib's method of communicating a potential violation of the DP guarantee. In their view, DP libraries should have stronger constraints that enforce privacy guarantees.
\begin{quoting}
Weird message, because I'm using a DP library to preserve privacy. And it's telling me there's this arbitrary privacy leakage. (D12)

\noindent So I just don't understand the choice at all. Why not just throw an error and not release the answer, and then require them to put in bounds on the data, or say we'll compute the bounds and in a way that will use up more privacy [budget]? (D9)
\end{quoting}
\noindent While Diffprivlib's warning handling was ambiguous in communicating potential privacy violations to users, OpenDP's error handling had a different issue: it often stymied participants, resulting in a long debugging process. The most common error message encountered by OpenDP users was a \emph{type mismatch} error. This error typically occurred when users were inconsistent in their use of integer and float types during a DP computation. While some participants were able to overcome this error, the frequency at which the error occurred led some participants to question the error's usefulness:
\begin{quoting}
Oh, again this same problem. This should be automatic. The library should infer the domain type. (O1)
\end{quoting}
OpenDP users also frequently encountered a \emph{dataset size} error, which happened when the size of the dataset was not specified when using OpenDP's DP mean function. Interestingly, participants with DP research experience actually \textit{welcomed} this error, as it reinforced some aspects of their mental model of DP libraries. 
\begin{quoting}
Okay this is good, because the size of the dataset is sometimes considered private and if it's not defined, then this could lead to a privacy violation (O2).
\end{quoting}
For some, encountering a dataset size error confirmed their understanding that the calculation of a DP mean may also require considering the size of the dataset. However, especially for participants without DP research experience, the amount of errors outputted by the library greatly increased user friction:
\begin{quoting}
At this point I'd have given up. Well, before this I probably would have given up. (O4)

\noindent I have no idea what this means. Can I check Stack Overflow? (O8)
\end{quoting}
\addition{Notably, OpenDP's errors conflicted with participants' initial mental models of DP libraries. For participants who were expecting to simply choose a function from the library that would make their output calculation differentially private (Fig \ref{subfig:b17model}), OpenDP's  frequent output of errors were often not expressive enough for participants to connect the necessary code change with the privacy violation or understanding mismatch that was occurring.}
\subsubsection{Libraries lacked consistent tooling to help users verify their DP results}
\label{sec:verify}
In order to simulate an instance of real-world data analysis task, we presented participants with two modified datasets: one with removed values (Dataset 2) and another with null values (Dataset 3). This task, described in Section \ref{sec:perturbation}, was designed to encourage participants to examine code implementations in the DP library and verify their own assumptions about library functions.

Notably, none of the participants assigned to Diffprivlib were able to notice the discrepancy in sensitivity calculation between the nanmean and mean functions after initial code completion, described in the previous section (Fig. \ref{fig:nanmean}). Only after they ran the given code that produced a histogram distribution of multiple DP mean computations, were they able to notice the difference in outputs.
\begin{quoting}
Oh, I'm very confused. Exactly why did we end up with 2 things that look so different? (D9)

\noindent I didn't expect that. But pretty interesting. (D14)
\end{quoting}
Participants gave several different reactions for the difference they saw. Most of them assumed that using the nanmean operation on Dataset 3 should have produced a similar output to using the mean function on Dataset 2.
\begin{quoting}
My initial reaction was that I thought that the Nanmean would just eliminate everything, and then it would treat it the same as Dataset 2. (D13)
\end{quoting}
On the other hand, some participants trusted the implementation of the nanmean function, even after seeing the discrepancy in outputs. Only two participants actually looked at the source code of the library in order to try and examine the implementation of the nanmean. Out of these two participants, only one of them, who was more experienced in DP, was able to discover the way in which Diffprivlib was calculating the sensitivity for the nanmean function:
\begin{quoting}
It seems wrong. Because they're just scaling sensitivity to the original size of the array, but dropping the NaNs. They're also directly using the number of NaNs, which itself is a private number. (D9)
\end{quoting}
While OpenDP did not have the same discrepancy in the calculation of the sensitivity of the mean function for the modified datasets, some participants noted a similar inability to verify the implementation of their code due to the difficulty of parsing through the documentation of OpenDP.
\begin{quoting}
The only way I could tell that my noise was set correctly was I just kept hitting run, and seeing if the numbers were sufficiently perturbed. (O5)
\end{quoting}
While A5's approach to verify their answer may have been safe to do in the context of the study environment, running the code execution multiple times is generally not a good practice, since it reduces the privacy guarantee of the computation. 

Relatedly, one of the main features of OpenDP's programming framework is its verifiability: if a computation is constructed using vetted components of the library, then a proof of privacy for that computation can be automatically derived \cite{OpenDP}. Currently, OpenDP has a \texttt{contrib} feature feature flag that users can enable to access constructors that have not been vetted. One participant suggested that OpenDP could do a better job in communicating the verified parts of the library to users.
\begin{quoting}
You could display the components of the computation and next to each part, there's a green check mark or lock that would show that the proof for each component is secure. You can do a lot as far as the communication of security without burdening people with the actual specificities. (O3)
\end{quoting}
Given that many of OpenDP's design choices are centered around ensuring that  measurements and transformations have their claimed privacy or stability properties, better communication of these concepts to users could help orient user mental models around the importance of using trusted and correct implementations.

Participants for both libraries also mentioned that having built-in visualization tools in DP libraries would have helped them verify the correctness of their answers.
\begin{quoting}
It could be nice to have some sort of visualization to show like the amount of noise that's being added. (D17)
\\
\noindent Graphs and visualizations are very useful in giving me the confidence to say what was predicted should have happened. (O7)
\end{quoting}
These findings across participants using both Diffprivlib and OpenDP suggest that there is a lack of tools available for participants who want to check the correctness of their implementations. In addition, even when tools such as OpenDP aim to reduce the burden on users by providing vetted code and modular components, they add complexity by requiring users to investigate frequent error messages and wade through dense documentation. In the next section, we discuss how implementation choices and different parts of the library relate to issues of trust and correctness.

\subsection{Building User Trust in DP Libraries}
In this final section, we discuss how DP libraries communicate notions of trust and correct implementation (RQ3). We find\addition{, based on qualitative questions about trust included in the semi-structured interviews,} that trustworthiness was signaled to users in a multitude of ways, including documentation, intuitive library design, and reliance on open-source library design. 
\addition{As trust in DP libraries is an emerging and mostly unexplored question, we did not have participants use quantitative measures of trust that are not-yet-validated in this new context. Instead, we provide initial, qualitative insights into the concept of trust in DP libraries and how trust is communicated to users, highlighting relevant factors that future work can further investigate.}

\begin{figure}[!htb]
\centering
\begin{subfigure}{0.4\textwidth}
    \includegraphics[width=\textwidth]{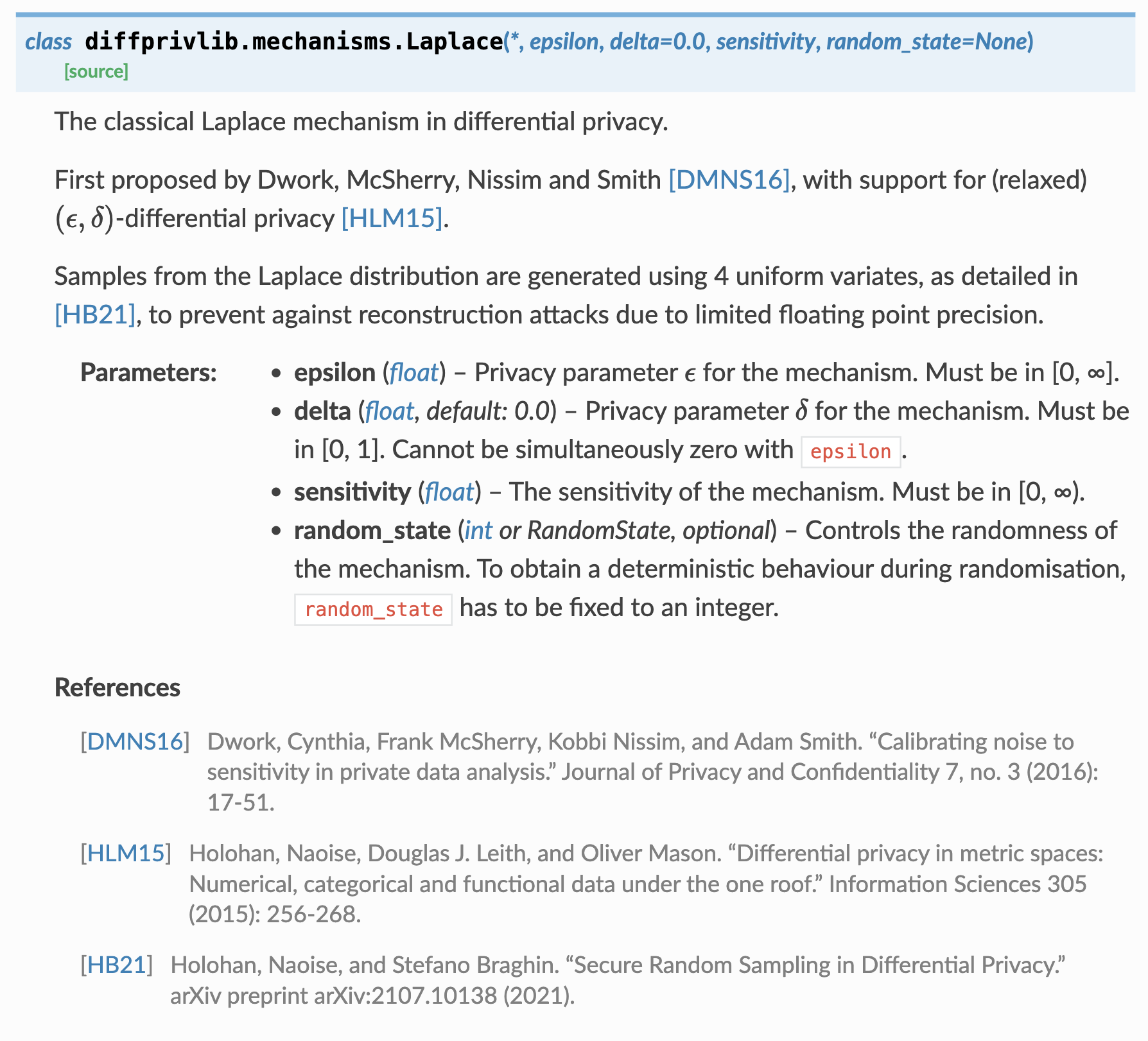}
    \caption{Diffprivlib}
\end{subfigure}
\begin{subfigure}{0.4\textwidth}
    \includegraphics[width=\textwidth]{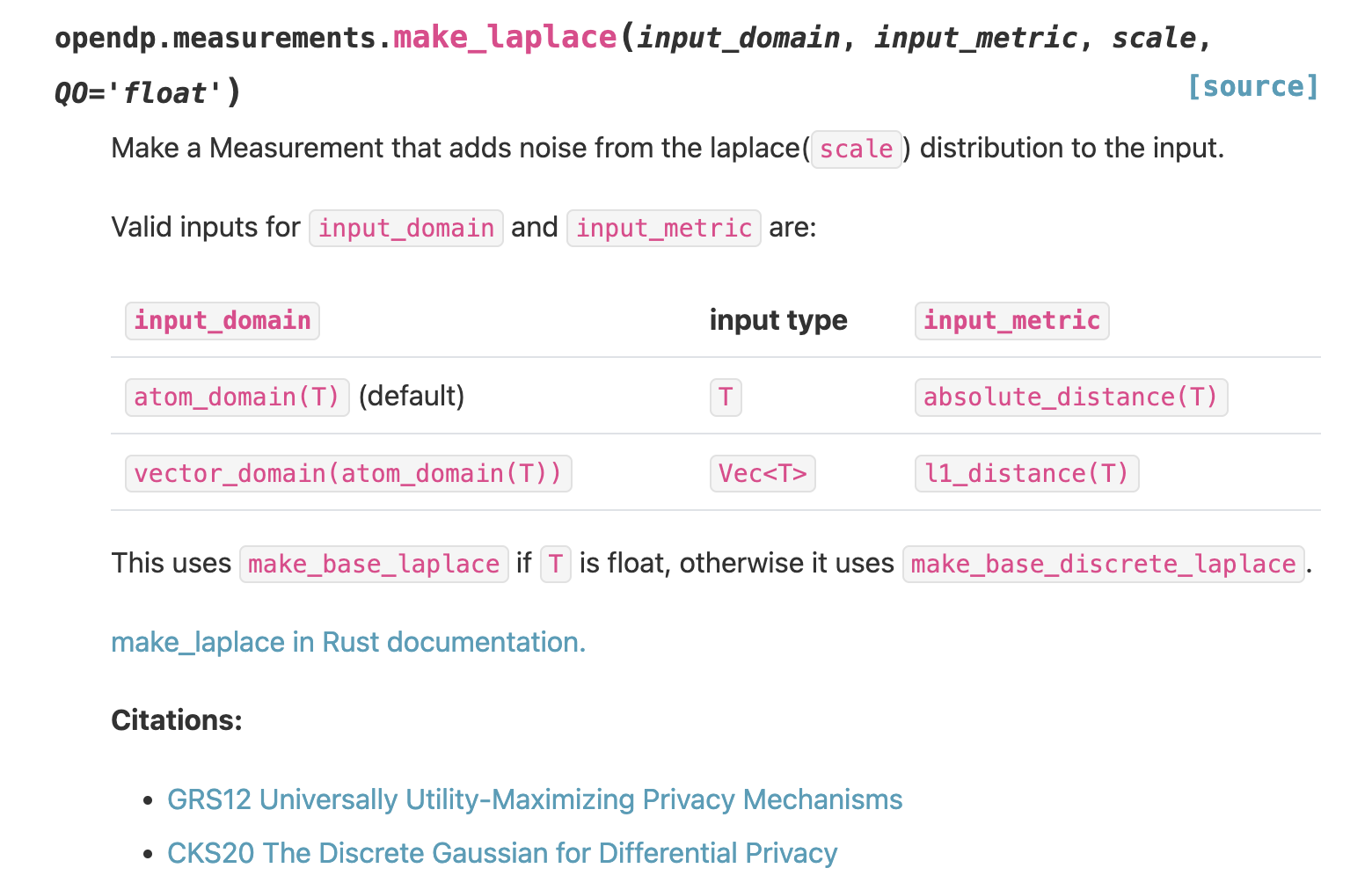}
    \caption{OpenDP}
\end{subfigure}     
\caption{Comparison of Diffprivlib and OpenDP documentation for the Laplace mechanism}
\label{fig:documentation}
\end{figure}

\subsubsection{Documentation is an important yet limited source of conveying trust}
\label{sec:documentation}
When asked questions about library trust, participants most frequently referred back to the choices the library made in designing the documentation. 
For some, library documentation communicated trust by citing outside sources. For example, linking research papers related to the source code of DP algorithms in libraries was one way libraries signalled correct implementation to users.
\begin{quoting}
I trusted it more because they link to a paper. So I knew they were pulling it from a legitimate source (O4).
\end{quoting}
Both OpenDP and Diffprivlib provided links to external research papers for their DP algorithms, when relevant (Fig. \ref{fig:documentation}). However, another participant noted that solely linking source code was not a sufficient remedy to overcome issues in the explainability of the documentation.
\begin{quoting}
Even if you open up the source code of these libraries. It's not super well explained what exactly those are and how they work. Sure, there might be external links to the papers that are being referred to. But I feel that they could be better explained (O2).
\end{quoting}
While some of the participants noted the presence of research paper citations in the documentation, none of them clicked on the external links to verify or further explore the relationship between the papers and the source code.

Interestingly, for two of the researcher participants, the ``Limitations'' page on OpenDP's documentation indicated a level of transparency that made them more comfortable about using the library. These participants expressed appreciation for the limitations page, which highlighted a number of known vulnerabilities in the library.
\begin{quoting}
Yeah, I think the limitations page is quite nice. It would make me feel better about using this library or at least trusting it to some level (O1).
\end{quoting}
For users who were more experienced in DP, the acknowledgement of limitations showed an awareness of difficulties in translating theoretical DP concepts to real-world implementations, and made them more willing to trust the library's implementations.

\subsubsection{Libraries balance maintaining an intuitive codebase with signalling correct implementation to users}
\label{sec:usabilitytrust}
Now, we return to the interaction between usability and privacy in DP libraries, a theme first explored in Section \ref{sec:tradeoff}. In Sections \ref{sec:models} and \ref{sec:programming}, we outlined how Diffprivlib's documentation and programming framework was generally easier for participants to understand, compared to that of OpenDP's. However, greater accessibility of the library's functions and code did not necessarily increase participant trust. One participant who uses libraries for experimentation noted:
\begin{quoting}
It's very well set up for someone like me, who's like just trying to test things rather than somebody who's actually trying to build something in a differentially private way (D9). 
\end{quoting}
For 4 participants who agreed to extend their user study session time and examine both OpenDP and Diffprivlib, we specifically asked them about their thoughts on how both libraries' treatments of usability and privacy guarantees affected their trust. When asked about this, B10 thought that OpenDP's relatively hard-to-use framework was necessary in order to enforce correct user behavior.
\begin{quoting}
You're dealing with people's data, right? Without that kind of steep learning curve, it's easy for users to make mistakes. (D10)
\end{quoting}
A7 believed that notions of trust is dependent on the specific audience the DP library is trying to reach. 
\begin{quoting}
If you're trying to target corporations who are releasing data publicly, I would recommend OpenDP. But for learning and investigation purposes, I would use Diffprivlib (O7).
\end{quoting}

For participants in both libraries, the mechanisms that mediated privacy budgets were singled out as being intuitive and useful. For example, A2 was happy to see that OpenDP's \texttt{Context} class properly restricted the number of queries they were able to ask, capping  to the number that was allocated in the initialization of the class.
\begin{quoting}
If I just specify split\_evenly\_over to be 1,  and I try to run 2 queries \ldots It's not gonna work. That's really good! (O2)
\end{quoting}
Similar restrictions were singled out for Diffprivlib as being beneficial for trust and utility. While B9 acknowledged some usability concerns with using Diffprivlib's budget accountant while in a code exploration stage, they noted its usefulness when using Diffprivlib for a real-world data release.
\begin{quoting}
It's a little unhelpful in the development process, because I think I'll run into issues where I accidentally exhaust the budget. But I think it would be super useful if I'm an analyst using this for an actual DP deployment. (D9)
\end{quoting}
These findings highlight the delicate balance that DP libraries strike between maintaining a user-friendly codebase, while also signalling to users that the implementations in the libraries are correct.

\subsubsection{Trust is multifactorial, relying on openness, proofs, and institutional contexts}
\label{sec:open-source}
Finally, we outline the multitude of ways in which participant perceptions of trust were affected during the DP library user study. For the most part, participants felt that the open-source nature of DP libraries helped them trust the library. The need to look at and verify the source code of the library was a frequent desire for participants.
\begin{quoting}
 I feel like I would like to read the code myself, and only if I trusted it, would I use it for any implementation. I don't think I can trust it in any other way (D10).
\end{quoting}
Additionally, two of the participants mentioned that having open-source code dissuaded any notion that these DP libraries were operating as a ``black-box`` system, where code implementations are exceedingly difficult to decipher.

Relatedly, an important part of OpenDP's process involves supplying mathematical proofs of the privacy properties of library functions. Participants had differing views on whether having privacy proofs in a DP library enhanced their trust of the library:
\begin{quoting}
I kind of inherently trust that the calculations [in the library] are being done correctly and it's not messing with anything fundamental. (D13)

\noindent You probably want to have those proofs somewhere, so that, a really committed person can verify the supposed security of the library. But the regular user doesn't need to understand them. (O3)
\end{quoting}
While providing privacy proofs is a priority from the perspective of library developers, the importance of having these proofs was not always communicated effectively to users.

Interestingly, some of the participants stated that institutional support of open-source software can improve their own trust of DP tools. Specifically, they pointed to OpenDP's backing by Harvard as a key factor for preferring open-source libraries over other proprietary software.
\begin{quoting}
 I really believe in open source. If it is backed by an institution like Harvard, I don't think there will be problems in implementing it in real situations (D11).

 \noindent For me, trust is built through knowing that the library has the consistent backing of the research community (O5).
\end{quoting}

In summary, findings from this section suggest that simply providing accessible documentation or vetted code does not, alone, engender trust in DP libraries. Rather, trust is built through a multitude of factors, from documentation and intuitiveness of code implementations, to support from specific institutions and organizations.

\section{Discussion and Future Work}
\label{sec:discussion}
Overall, we find that while open-source libraries for DP are important and promising tools for widening access to and trust in privacy-preserving data analysis, \addition{developing software for DP poses unique challenges regarding trust and usability. Our study illustrates how} participants encountered significant gaps between conceptual and mental models of DP libraries, \addition{and shows how} these gaps are not sufficiently bridged by the libraries' open-source code, documentation and error messaging. Furthermore, our findings highlight the the diverse ways in which \addition{DP} libraries can communicate trust and safety to users. 

Our analysis of open-source libraries for DP offers insights for building trustworthy data pipelines \addition{in general}, such as taking a multidimensional, socially situated approach to building trust and recognizing the ways in which rigour and usability are entangled. Yet, perhaps the most important takeaway is that the concept of usability itself is different, and more complex, for DP compared to other types of data analysis tools, something that the developers in our study already grapple with but that requires further attention from the broader DP and HCI communities. Thus, we highlight below the specificities of DP regarding design, trust, and usability and outline practical suggestions for DP library development.

\subsection{Lessons from analyzing open-source DP libraries}

Our goal in studying open-source DP libraries is twofold: (1) analyzing and improving tools that can lead to rigorous privacy protections across a wide range of data applications, and (2) shedding light on the challenges of designing tools for responsible privacy software more broadly.

In particular, our study highlights the need to think deeply about trust, transparency, usability, and accountability beyond what is provided via technical means such as open-source code and theoretical, algorithmic guarantees. 

\subsubsection{Trust and transparency starts, but doesn't end, with open-source code} 

Our study highlights the fallacy of equating open-source code with transparent and usable \addition{DP} software. Even after spending time reading documentation and writing functions, participants---even those with experience in DP---exhibited significant misunderstandings of how to use the software. While the nature of these misunderstandings differed between OpenDP and DiffPrivLib, both libraries failed to orient users towards safely and effectively using DP.

This leads to the question: what is the role of open-source libraries in empowering data analysts and enabling new deployments of DP? Our study corroborates the findings from other \addition{DP-focused} studies that open-source software is an important step towards creating transparent, usable, and trustworthy DP deployments but is not enough on its own~\cite{ngong23,Smart22,Dwork2019}. Participants in our study emphasized, in particular, that trust in DP software is built through a variety of factors beyond simply having open-source code and mathematical proofs, such as: designing programmatic frameworks to match both the theoretical literature and/or data science tools, crafting effective explanations and error messages, and having trustworthy institutional backing and expertise. 
\addition{These results further open up questions around trust in DP software, providing initial qualitative insights that can serve as a foundation for future quantitative studies. Building on these factors our study has identified as relevant for trust, future work can use approaches such as conjoint analyses (e.g. ~\cite{ayalon2023exploring}) to provide generalizable insights into trust in DP software.}
\removal{As such,} \addition{In addition,} those involved in bringing DP from theory to practice should keep in mind that trust and transparency, like privacy itself, are social concepts~\cite{hissam2002trust,ho2013trust,Smart22}, and that users’ motivations for (dis)trusting DP may be broader than having trusted code or provable privacy guarantees.

\subsubsection{Rigour and \addition{friction-less interaction} may seem to be in tension, but \addition{are both important for usable DP}}
%can also be mutually supportive for DP .}
Indeed, our study highlights that a key challenge for designing usable DP tools is the tension between \removal{rigour and usability} \addition{frictions that enhance rigour and frictions that hamper user interaction}. In particular, we find that while Diffprivlib attempts to cater to a broad audience by aligning with existing data-science libraries, its \addition{low-friction} programmatic approach may lead to frequent privacy violations by users \addition{and thus, less usability}. In contrast, OpenDP's adherence to rigorous privacy guarantees and implementations can cause \addition{undue frictions and again create} usability hurdles \addition{in the opposite direction} for data analysts. 

Concurrent work on open-source DP libraries~\cite{ngong23} points to a middle-ground approach taken by Tumult. Ngong et al. found that Tumult Analytics~\cite{Hay22}, a DP library for computing aggregate queries on tabular data.
On the one hand, like Diffprivlib, Tumult Analytics models itself to some extent on existing tabular data libraries like Pandas and Spark. On the other hand, the Tumult analytics library is built upon a privacy foundation library that provides some of the same types of rigorous guarantees offered by OpenDP. So, it seems that a middle ground is possible. However, we note that the computations provided by Tumult are limited to tabular data, while Diffprivlib and OpenDP expose a wider range of functionality for different data types. Taking this middle ground may be challenging when designing a library for a broad audience and set of use-cases. 
\addition{Future work that expands the inquiry of our study to Tumult and other tools in the growing landscape of usable DP should interrogate these additional tradeoffs around usability.}

Our study also suggests that while rigour and \removal{usability}\addition{ease-of-use} might seem to be on opposite sides of a spectrum, there are ways in which each one is supportive of the other. As we heard from both developers and researchers, rigorous design of the library can enable users to build more complex DP mechanisms, while \removal{usable}\addition{easy-to-use} code is important for users to understand and maintain the guarantees offered by DP.
Therefore, it may be more productive to think about ways in which to enhance rigour and \removal{usability}\addition{reduce frictions} \emph{in tandem}. Our study suggests that documentation and error messaging are powerful tools to align conceptual and mental models towards these goals, and we expand on this more in Section~\ref{sec:recommendations}. 

\iffalse
\subsubsection{The impact of design choices on trust relationships should be clarified.}
Finally, our study raises the question of accountability. When making decisions about rigour and usability, who holds the burden of accountability when something goes wrong?
Through comparing OpenDP and Diffprivlib, our study highlights that different design choices yield different trust relationships, placing burdens of accountability on different parties at different times. For example, OpenDP places a greater burden on the user to understand the programmatic framework and constraints of DP before deployment, while Diffprivlib implicitly requires trusting users to not use the software in production and to debug their own privacy errors in real-time. Each approach has its merits in different contexts, yet far too little attention has been placed on how design choices encode these trust relationships. Future work should aim to clarify these assumptions and relationships, as well as make clear how each design choice can be paired with governance, communication, and educational mechanisms that can mitigate misuse of DP tools.
\fi

\subsection{Practical recommendations for DP library development}
\label{sec:recommendations}

Through insights derived from the formative interviews and DP library user study, we present a set of recommendations aimed at enhancing user experience and safety when interacting with DP libraries. These recommendations are informed by the challenges and needs identified by users and developers in understanding and effectively implementing DP concepts. We focus on four key areas: error messaging, language and communication, use of visualizations, and establishing stronger norms when handling data. 
\begin{enumerate}
    \item \textit{Use \removal{more} expressive error messaging \addition{that better communicates unintuitive, DP-specific privacy violations}}. Through the user study, we found that error and exception messaging was one of the main ways that libraries interact with users to orient user mental models to better align with the library's conceptual model. \addition{While this is true for
    all software libraries, such alignment is particularly critical for DP libraries, in order to ensure not just ease-of-use but also correctness and safety. In both libraries, the error messaging was not expressive enough to create such alignments;} for example, participants were able to infer that OpenDP's errors had to do with some incorrect syntactic choice, but they weren't able to contextualize their error in terms of the exact privacy violation that was occurring. \addition{And when using Diffprivlib, many participants were not able to understand the impact of using incorrect bounds and how this might be a DP violation of from just the sparse warning given by the library.}
    Since the community surrounding open-source DP libraries is not as extensive as other data science libraries, the effectiveness of user troubleshooting through online resources is limited. Thus, DP libraries should better anticipate common user mistakes, and use error messaging and documentation as mechanisms to develop user knowledge of DP. 

    In Figure \ref{fig:modifiedmessages}, we propose a modified error message \addition{to be shown to users} when they leave the size and bounds parameters of a DP mean function unspecified. This error message allows users to use part of their privacy budget to privately compute the size and bounds parameters, and explicitly suggests a code change that would allow them to fix the error. It also links to additional documentation which outlines best practices for determining dataset attributes safely. Additionally, \addition{we recommend} having a user flag that can adjust the expressiveness of error messages based on the \addition{DP} expertise of the user to better accommodate \removal{the use cases of} both beginners and experts.

\begin{figure*}[t!]
    \centering
    \begin{subfigure}[t]{0.48\textwidth}
        \centering
        \includegraphics[width=\textwidth]{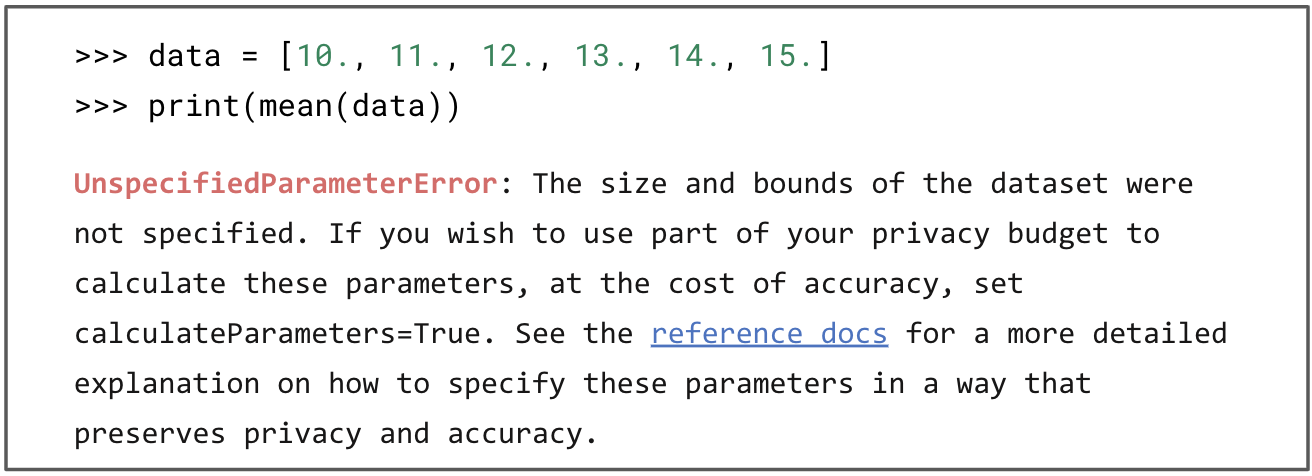}
        \caption{Unspecified parameter error}
    \end{subfigure}
    \hspace{5mm}%
    \begin{subfigure}[t]{0.48\textwidth}
        \centering
        \includegraphics[width=\textwidth]{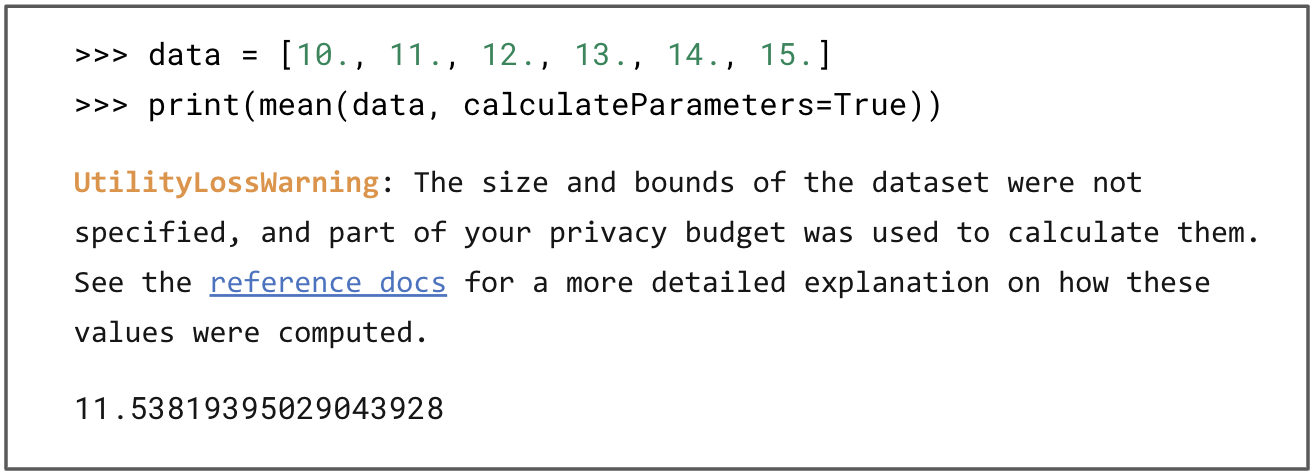}
        \caption{Bounds and size warning}
    \end{subfigure}
    \caption{Proposed library messages}
    \label{fig:modifiedmessages}
\end{figure*}

    \item \textit{Design libraries to speak the language of \addition{intended end users of DP by highlighting accuracy, in addition to privacy.}} Given that many aspects of DP are hard to understand for users, from choosing an optimal value for the privacy-loss parameter \cite{Dwork2019, Lee11} to contextualizing the error of a DP release \cite{cummings21, Franzen22}, we recommend communicating and educating users on \addition{making informed DP choices} \removal{proper privacy choices} through more familiar domains. As suggested by the developer interviews from Section \ref{sec:formativeobservations} and the responses from Section \ref{sec:errors}, the approach taken by Diffprivlib in inferring dataset bounds is ineffective towards orienting users towards correct privacy-preserving DP practices. While Diffprivlib's approach in outputting a warning instead of an error is conducive towards decreasing user friction, it doesn't clearly communicate the consequences for not specifying any dataset bounds \addition{with respect to the DP guarantee}. Instead \addition{of focusing just on privacy implications}, we recommend directing users \removal{to specify sensitivity-affecting parameters}\addition{towards correct practices around setting bounds} by communicating the consequences of their actions \removal{through utility loss}\addition{in terms of loss in accuracy} (as shown our proposed error message in Figure \ref{fig:modifiedmessages}), which is a domain that is more familiar to users. This approach communicates to users that proper \addition{selection of bounds and parameters} \removal{choices} can lead to increased accuracy, and can also better \removal{educate and} motivate users to understand how the privacy-accuracy tradeoff can be improved.

    \item \textit{\addition{Incorporate visualizations to convey the probabilistic nature of DP}}. When participants were asked about ways in which correct implementation can be communicated to them, visualizations were one of the most mentioned responses. However, in our formative interviews, incorporating more visualization tools was not mentioned by any of the developers in the formative interviews as a focus for future library development. Given the development of effective visualization tools to better help users make informed choices about DP \cite{Panavas23, Xiao18, nanayakkara22}, we suggest incorporating more visualizations in the documentation and example notebooks of DP libraries. \addition{In addition, we recommend having built-in functions and graphic visuals in the library, such as the frequency-framed visualizations described by Nanayakkara et al. \cite{Nanayakkara23}, that can better convey and contextualize the probabilistic nature of DP operations.} This would reduce the responsibility of library users to create their own visualization methods.

    \item \textit{Establish norms \addition{around data analysis that align with the practical constraints of DP:} Don't look at the data!} 
    As detailed by Sarathy et al. \cite{Sarathy23}, DP fundamentally changes every step of the data science workflow. One of the most challenging changes to the normal data science workflow is the practice of looking at the raw dataset. For some participants in \removal{the}\addition{our} user study, one of the first steps they took in the data analysis was to look at the data, either by directly examining the csv file or by viewing a subset of the dataset via a print operation. Others refused to look at the data, and relied on the metadata provided in the task prompt to choose parameter values for their DP functions. Given that looking at the data can both directly and indirectly lead to privacy violations, we recommend that DP libraries establish a norm that users should not look at the underlying dataset. Through documentation and example notebooks, libraries can better explain how looking at the dataset can reduce the effectiveness of privacy guarantees and encourage processes in which sensitivity-affecting parameters can be safely calculated or inferred from the dataset or prior knowledge. In order to better facilitate library use for beginners, we recommend that DP libraries either implement synthetic data generation functions or include resources  that allow users to create non-sensitive versions of their datasets. Doing so would help guide users towards \addition{DP-compatible} data practices that encourage exploration on non-sensitive datasets, but limit exploration on sensitive ones.
\end{enumerate}

% \subsection{Limitations}
% As open-source DP libraries are emerging tools, this study is \removal{meant to be} an initial, \additon{qualitative} exploration. It does not consider all the available DP libraries---and chooses only two for participants to interact with. \emphalthough analyzing these libraries can shed light on the others, and on open-source tools for responsible AI more broadly. Further work is needed to understand

\section{Conclusion}
\label{sec:conclusion}
In this study, we analyze the differences between conceptual models embodied in open-source DP libraries and mental models of DP held by users. Through interviews with 5 developers of open-source DP libraries and
user studies with 17 data analysts, we investigate the gaps between theoretical foundations and implementation-level challenges of DP, as well as how DP libraries communicate trustworthiness and correct implementation. We find that library usability does not always facilitate better understanding of DP concepts, and that DP libraries often struggle to guide users towards correct implementations. In addition, our work highlights the multifaceted role of trust within DP libraries.

Future work should extend our study of two libraries to the broader space of DP libraries and tools.
The insights gained from this study show that the design choices of DP libraries greatly affect how users understand, perceive, and trust DP. The findings also highlight the versatility in using mental models as a framework to encode and compare understandings of DP. We hope that our work provides guidance towards future development of usable, safe DP tools.

\section{Acknowledgements}
This research work was funded by the Alfred P. Sloan Foundation (Award No. G-2022-19501), National Science Foundation (Award No. TI-2303681), and the Harvard Faculty Aide program.

\bibliographystyle{ACM-Reference-Format}
\bibliography{acmart-primary/acmart}

\newpage

\appendix

\section{Mental Model Drawing Task}
\label{sec:mental-model-drawing}
Participants will be asked to complete the following drawing exercise. The prompt is described as follows:
\textit{This is a drawing exercise. I’m going to ask you to draw how you think the open-source differential privacy libraries work in practice. Please talk aloud and explain your thought processes while you are drawing. Please keep in mind that there is no correct answer to these questions—just answer these questions based on your own knowledge and experiences.}

\noindent \textbf{Developer Questions}
\begin{enumerate}
    \item Choosing any example of a computation on data (summary statistic, statistical model, algorithm, etc.), can you draw and explain how your library implements a DP computation and execute it on a sensitive dataset.
    \item Highlight the parts of the library that are most secure/vulnerable to privacy attacks or leaks. 
    \item In working within your library, when/where are users most responsible (or not responsible) for ensuring a correct, differentially private output? In what parts are the library developers responsible (or not responsible) for ensuring a correct, differentially private output?
    \item What parts of your library are important to directly communicate with users? What parts can be abstracted from the user? (answered with tumult core vs tumult analytics distinction)
    \item How do libraries communicate notions of trust and security to users? 

\end{enumerate}
\textbf{User Study Questions}
\begin{enumerate}
    \item Choosing any example of a computation on data (summary statistic, statistical model, algorithm, etc.), can you draw and explain how you think a DP library would implement a DP computation and execute it on a sensitive dataset?
    \item Please highlight the parts of the library that you perceive to be most secure/vulnerable to privacy attacks or leaks. 
    \item What role do DP libraries play in communicating trust during the mental model diagram exercise?
    \item Where are you (the user) responsible for ensuring that there is a correct DP output? Where is the library responsible for ensuring that there is a correct DP output? 
\end{enumerate}

\section{Library Developer Interview Questions}
\label{sec:developer-questions}
\begin{enumerate}
    \item What is your background with differential privacy?
    \item Who is the primary intended user(s) (researcher, data scientist, etc.) for your DP library?
    \item What design choices or considerations are important in your experience when implementing open-source differential privacy algorithms?
    \item What are the biggest obstacles you have noticed for users of your library?
    \item What insights have you learned from developing open-source DP software that can be applied to creating future DP tools?
    \item What is different about your library compared to other libraries?
    \item How well do practices and design choices in your library generalize to other tools and domains? Are there any practices and/or design choices that are easily adoptable for developers and practitioners?
\end{enumerate}

\section{User Study Task Instructions}
\label{sec:user-study-instructions}
\textbf{Task 1: DP Mean}
\begin{enumerate}
    \item Examine the dataset\_1.csv file. Comment on the information that the file conveys.
    \item Using functions from OpenDP/Diffprivlib, define a function that computes the DP mean of the income values in dataset\_1.csv. Use a privacy budget of $\varepsilon$=1, and assume that the income values range between 0 and 300,000. Feel free to use the documentation given above, or to explore documentation from other parts of the library.
    \item Using the compute\_histogram function defined in the notebook, construct and analyze the distribution of DP mean releases
    \item Now, imagine a scenario where individuals earning over 45,000 do not reveal their incomes. Examine the dataset\_2.csv and dataset\_3.csv files. What similarities and differences do you notice between these two datasets? What do you predict the DP mean income values will be for these datasets?
    \item Repeat steps 1 and 2 using the dataset\_2.csv and dataset\_3.csv files. 
    \item Compare the histogram releases that you computed for all 3 datasets. Do the results align with what you predicted?
\end{enumerate}
\textbf{Task 2: Histograms and Budget Accountant}
\begin{enumerate}
    \item First, set up a privacy budget accountant with a total privacy budget of $\varepsilon$=1
    \item Now, using dataset\_1.csv, compute a DP histogram of the ages in the dataset. Use 10 bins for the histogram, and an epsilon value of 0.8.
    \item Now, compute a DP histogram using an epsilon value of 0.25. 
    \item Compare the two histograms that you computed above. Did result of changing the epsilon value align with expectation?
    \item Now, check how much privacy loss you have incurred so far and the remaining privacy budget you have.
\end{enumerate}
\textbf{Post-task Questions}
\begin{enumerate}
    \item Returning to the mental model diagram you drew at the beginning of the study, is there anything that you would change about the diagram you drew?
    \item Are there any changes you would make to the documentation of the library you were examining?
    \item What is the value of open-source software in creating an environment that is conducive to DP development?
\end{enumerate}

\newpage
\section{Codebook}
\label{sec:codebook}
\begin{center}
\newcolumntype{s}{>{\hsize=.5\hsize}X}
\renewcommand{\arraystretch}{1.25}
\begin{table}[htp]
  \begin{tabularx}{\textwidth}{p{0.45\textwidth}p{0.45\textwidth}}
    \FL
 \textbf{Theme} / Code
    \ML
    \textbf{Implementing DP algorithms} \newline
    \begin{itemize}
        \item Using DP libraries for deployments
        \item Importance of correct implementation
        \item Practical implementation of theoretical research
    \end{itemize} 
    & 
    \textbf{Library Usability} \newline
    \begin{itemize}
        \item Usefulness of visualizations
        \item Explainable DP code
        \item Misunderstanding fundamental DP library norms
    \end{itemize}
    \ML
    \textbf{Open-source software (OSS)} \newline
    \begin{itemize}
        \item Institutional backing of DP OSS
        \item Contributing to DP OSS
        \item Helpfulness of transparent algorithms
    \end{itemize}
    & 
    \textbf{Trusting DP Libraries} \newline
    \begin{itemize}
        \item Communicating trust to users
        \item Vulnerabilities of DP libraries
        \item Expectations of library safeguards
    \end{itemize}
    \ML
    \textbf{Programming in OpenDP} \newline
    \begin{itemize}
        \item Considering bounds before computing mean
        \item Defining domain
        \item Naming mismatch
        \item Privacy unit and loss
        \item Measurements and transformations
        \item Splitting privacy budget
        \item Documentation confusion
    \end{itemize} 
    & 
    \textbf{Programming in Diffprivlib} \newline
    \begin{itemize}
        \item Reasoning about the bounds
        \item Bounds warning
        \item Handling null values
        \item Calculating nanmean
        \item Usefulness of privacy accountant
        \item Intuitive documentation
    \end{itemize}
    \LL
  \end{tabularx}
  \vspace{10pt}
    \caption{Summary of research themes and codes}
    \label{table:codebook}
\end{table}
\end{center}

\newpage
\section{Additional Figures}
\label{sec:additional-figures}

\begin{figure}[!htb]
    \centering
    \includegraphics[width=0.50\linewidth]{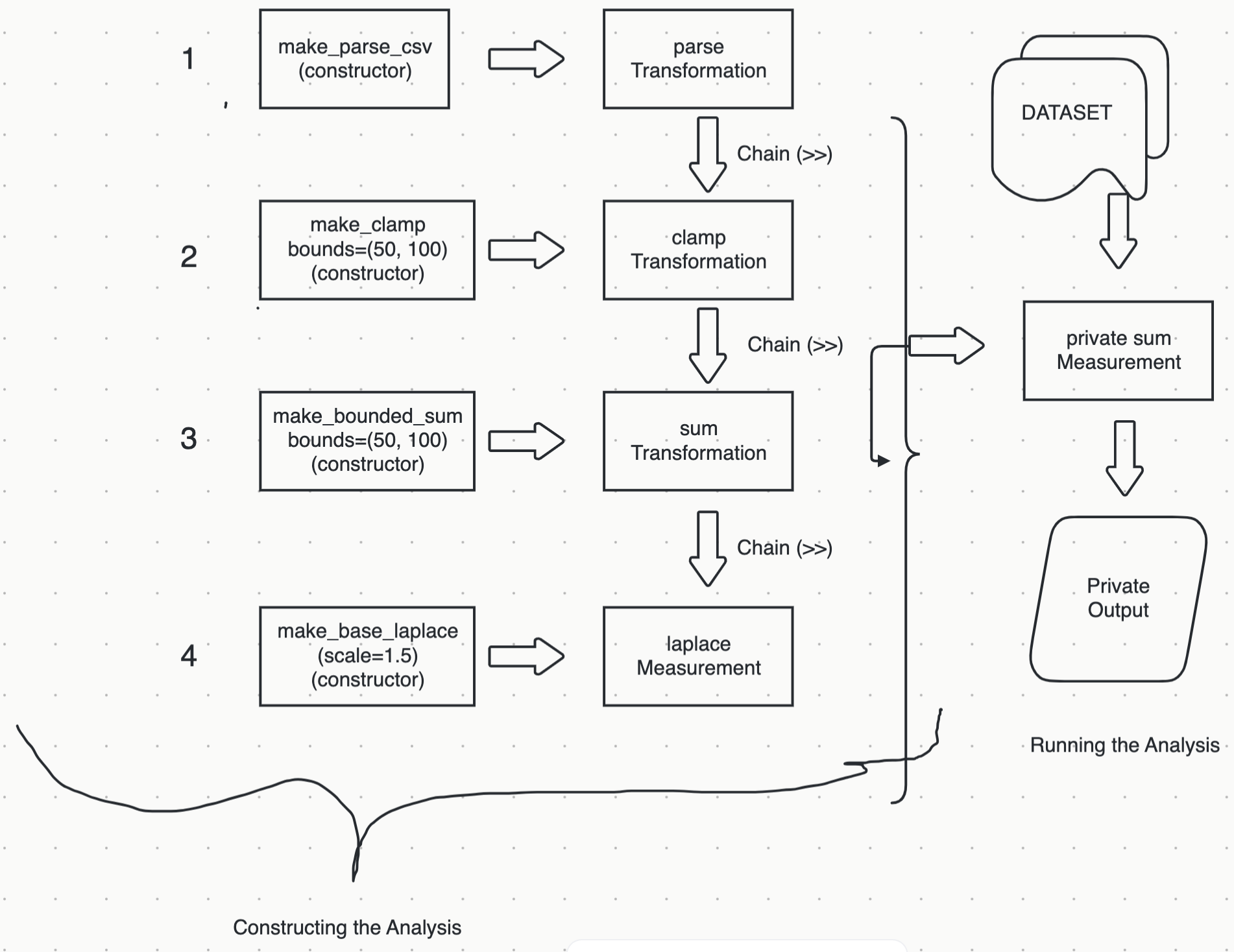}
    \caption{Library Conceptual Model (P3)}
    \label{fig:devmodel}
\end{figure}

\begin{figure}[!htb]
    \centering
    \includegraphics[width=0.85\linewidth]{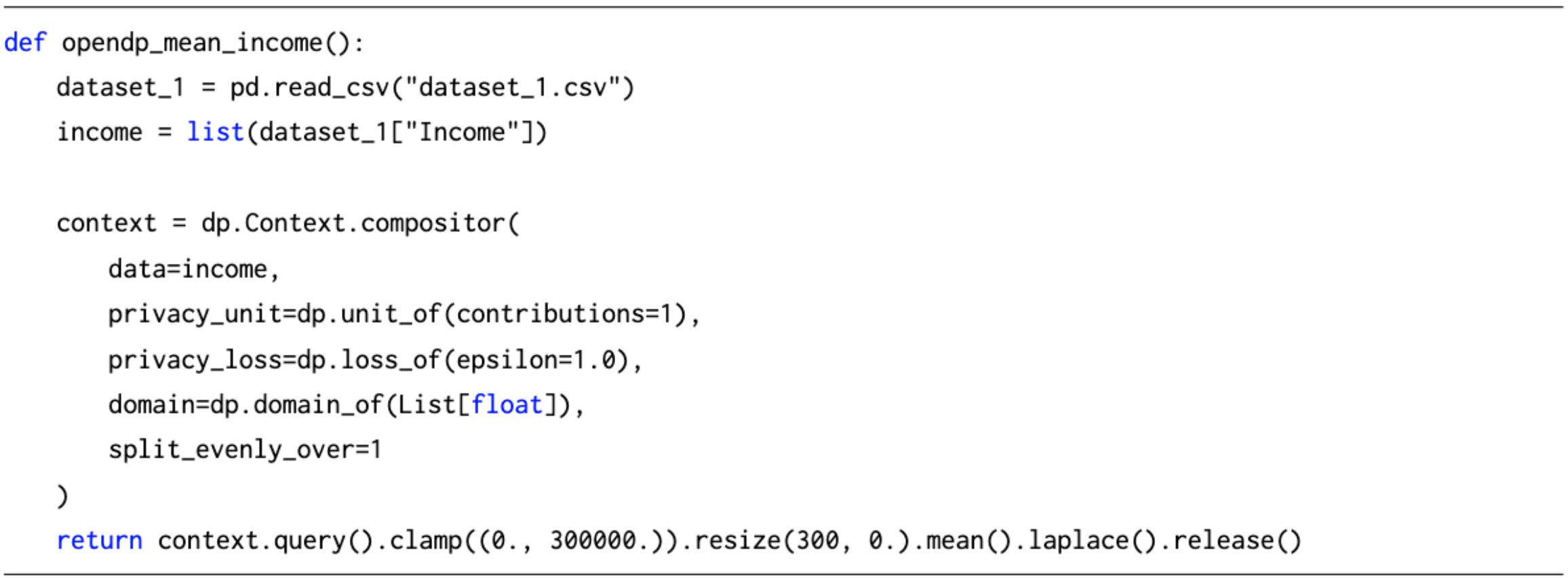}
    \caption{OpenDP Mean}
    \label{fig:opendpmean}
\end{figure}

\begin{figure}[!htb]
    \centering
    \includegraphics[width=0.85\linewidth]{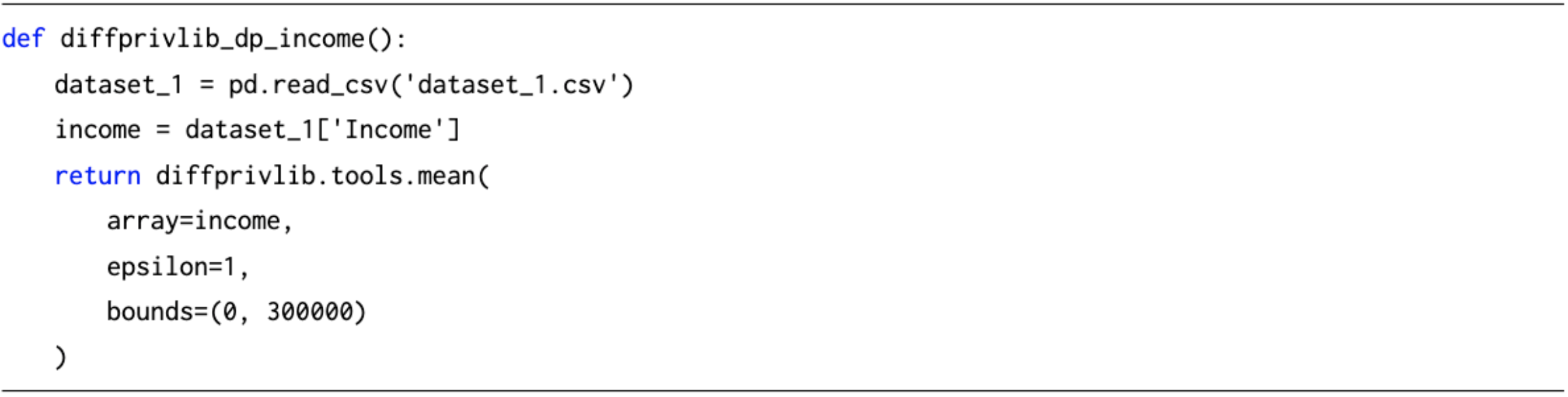}
    \caption{Diffprivlib Mean}
    \label{fig:diffprivlibmean}
\end{figure}

\begin{figure}[!htb]
     \centering
     \begin{subfigure}{\textwidth}
         \centering
         \includegraphics[width=0.65\textwidth]{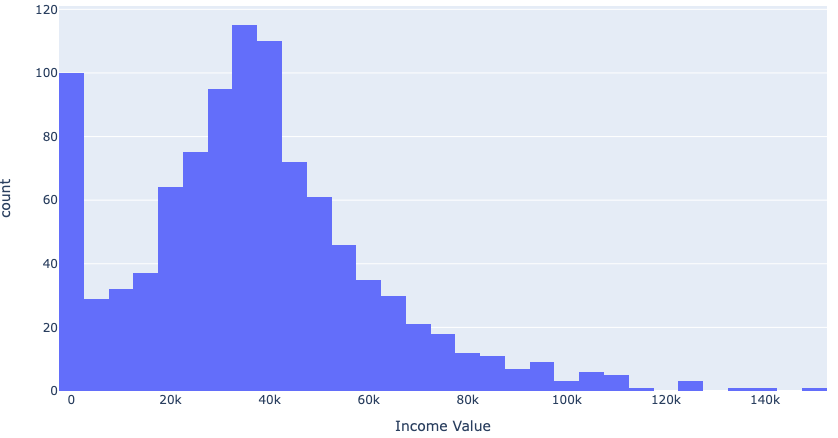}
         \caption{Distribution of 1000 DP mean computations on dataset with removed values}
     \end{subfigure}
     \hfill
     \begin{subfigure}{\textwidth}
         \centering
         \includegraphics[width=0.65\textwidth]{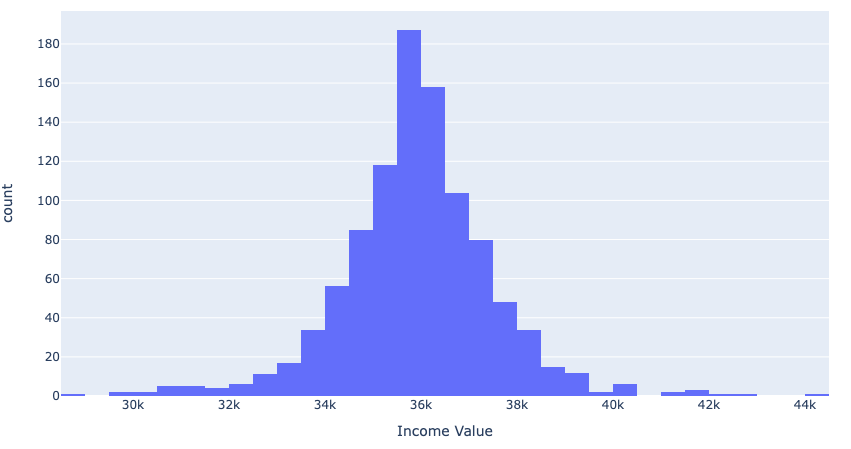}
         \caption{Distribution of 1000 DP nanmean computations on dataset with NaNs}
     \end{subfigure}
     \hfill
     \caption{Comparison of Diffprivlib DP mean and DP nanmean functions}
     \label{fig:nanmean}
\end{figure}

\begin{figure}[!htb]
     \centering
     \begin{subfigure}{\textwidth}
         \centering
         \includegraphics[width=0.85\textwidth]{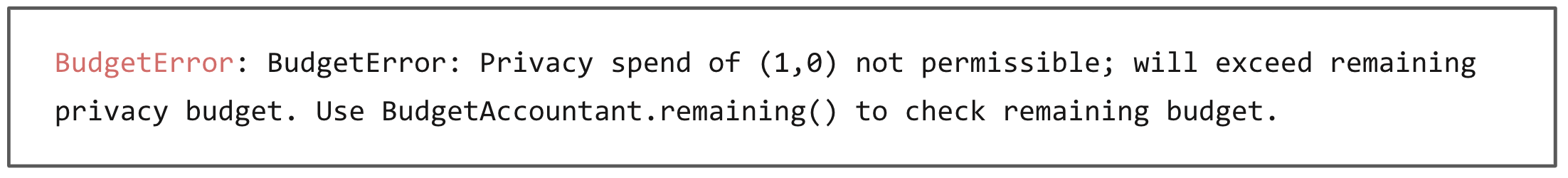}
         \caption{Diffprivlib Budget Error}
     \end{subfigure}
     \hfill
     \begin{subfigure}{\textwidth}
         \centering
         \includegraphics[width=0.85\textwidth]{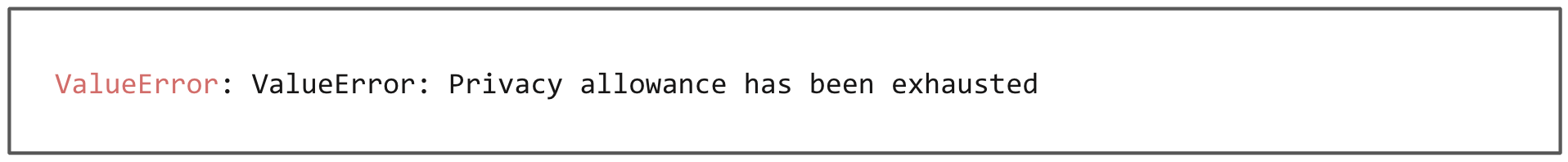}
         \caption{OpenDP Budget Error}
     \end{subfigure}
     \hfill
     \caption{Comparison of Diffprivlib and OpenDP privacy budget exceeded error messages.}
     \label{fig:budget}
\end{figure}

% \begin{figure}[!htb]
%     \centering
%     \includegraphics[width=0.5\linewidth]{figures/articulated_model_1.png}
%     \caption{Bidirectional Complex Model (D10)}
%     \label{fig:b10model}
% \end{figure}

\begin{figure}[!htb]
    \centering
    \includegraphics[width=0.5\linewidth]{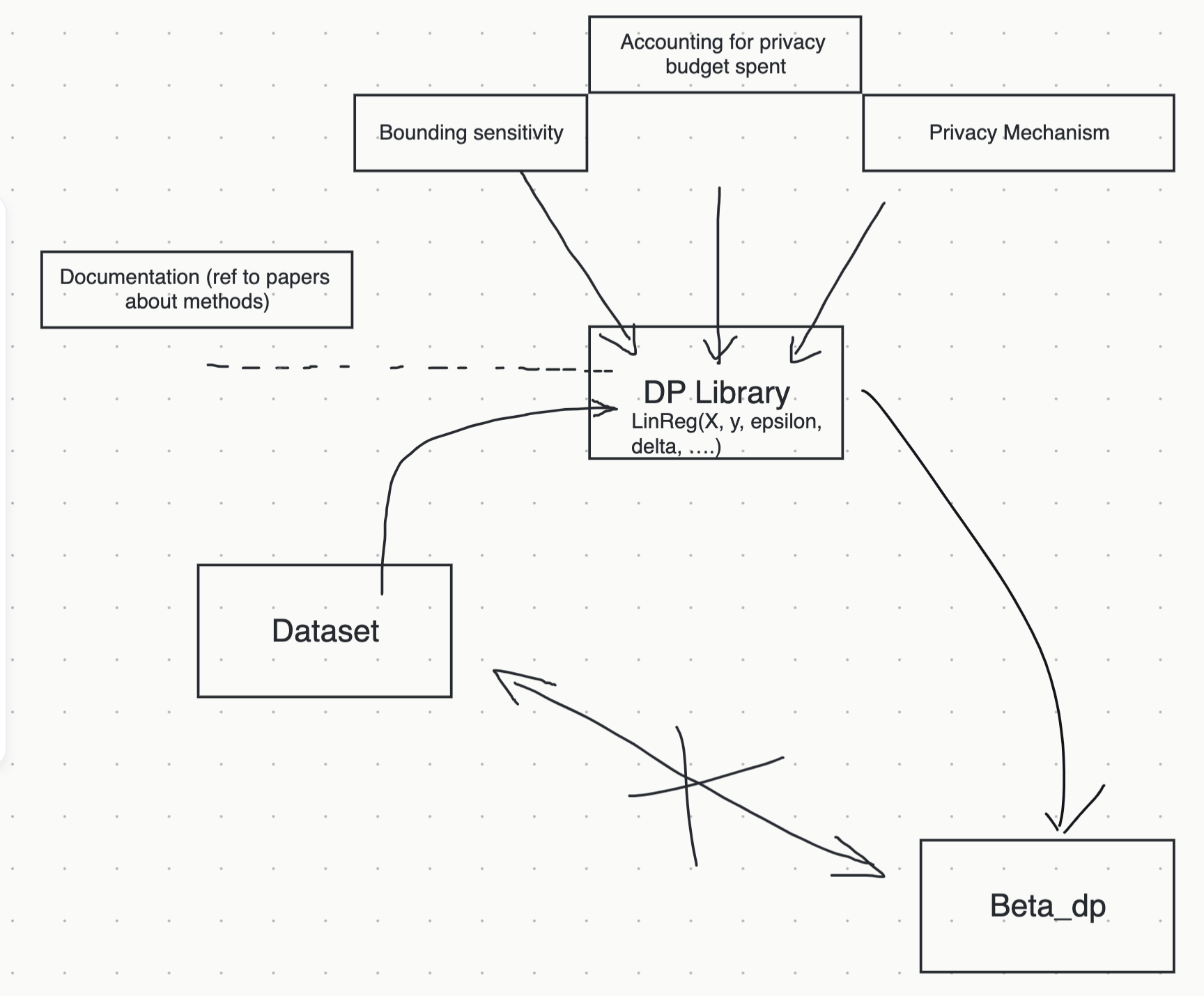}
    \caption{DP Linear Regression Model (O1)}
    \label{fig:A1model}
\end{figure}

% \begin{figure}[!htb]
%     \centering
%     \includegraphics[width=0.5\linewidth]{figures/simple_model_1.png}
%     \caption{Model of Utility Cost Computation (D17)}
%     \label{fig:b17model}
% \end{figure}

\begin{figure}[!htb]
    \centering
    \includegraphics[width=0.7\linewidth]{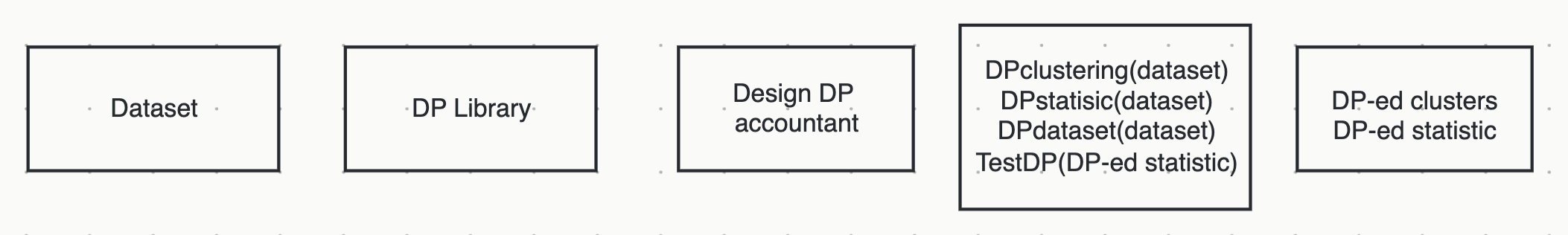}
    \caption{Model of DP Clustering (D14)}
    \label{fig:b14model}
\end{figure}
\newpage

\end{document}